\documentclass[]{emulateapj}
\usepackage{apjfonts}
\usepackage{mathptmx}

\slugcomment{To be published in {\it The Astronomical Journal}.}

\shorttitle{MULTIBAND IMAGES OF DISK AROUND $\beta$ PICTORIS}
\shortauthors{GOLIMOWSKI ET AL.}
 
\begin{document}
 
\title{{\it HST}/ACS Multiband Coronagraphic Imaging of the Debris Disk around Beta Pictoris\altaffilmark{1}}

\author{
D.~A.~Golimowski,\altaffilmark{2}
D.~R.~Ardila,\altaffilmark{3}
J.~E.~Krist,\altaffilmark{4}
M.~Clampin,\altaffilmark{5}
H.~C.~Ford,\altaffilmark{2}
G.~D.~Illingworth,\altaffilmark{6}
F.~Bartko,\altaffilmark{7}
N.~Ben\'{\i}tez,\altaffilmark{8}
J.~P.~Blakeslee,\altaffilmark{9}
R.~J.~Bouwens,\altaffilmark{6}
L.~D.~Bradley,\altaffilmark{2}
T.~J.~Broadhurst,\altaffilmark{10}
R.~A.~Brown,\altaffilmark{11}
C.~J.~Burrows,\altaffilmark{12}
E.~S.~Cheng,\altaffilmark{13}
N.~J.~G.~Cross,\altaffilmark{14}
R.~Demarco,\altaffilmark{2}
P.~D.~Feldman,\altaffilmark{2}
M.~Franx,\altaffilmark{15}
T.~Goto,\altaffilmark{16}
C.~Gronwall,\altaffilmark{17}
G.~F.~Hartig,\altaffilmark{11}
B.~P.~Holden,\altaffilmark{6}
N.~L.~Homeier,\altaffilmark{2}
L.~Infante,\altaffilmark{18}
M.~J.~Jee,\altaffilmark{2}
R.~A.~Kimble,\altaffilmark{5}
M.~P.~Lesser,\altaffilmark{19}
A.~R.~Martel,\altaffilmark{2}
S.~Mei,\altaffilmark{2}
F.~Menanteau,\altaffilmark{2}
G.~R.~Meurer,\altaffilmark{2}
G.~K.~Miley,\altaffilmark{15}
V.~Motta,\altaffilmark{18}
M.~Postman,\altaffilmark{11}
P.~Rosati,\altaffilmark{20}
M.~Sirianni,\altaffilmark{11}
W.~B.~Sparks,\altaffilmark{11} 
H.~D.~Tran,\altaffilmark{21}
Z.~I.~Tsvetanov,\altaffilmark{2}   
R.~L.~White,\altaffilmark{11}
W.~Zheng,\altaffilmark{2}
and
A.~W.~Zirm\altaffilmark{2}
}

\altaffiltext{1}{ 
Based on guaranteed observing time awarded by NASA to the ACS Investigation Definition Team ({\it HST} program 9987).
}
\altaffiltext{2}{ 
Department of Physics and Astronomy, 
The Johns Hopkins University, 
3400 North Charles Street, 
Baltimore, MD 21218-2686
}
\altaffiltext{3}{ 
Spitzer Science Center,
Infrared Processing and Analysis Center,
MS 220-6,
California Institute of Technology,
Pasadena, CA 91125
}
\altaffiltext{4}{ 
Jet Propulsion Laboratory,
4800 Oak Grove Drive,
M/S 183-900,
Pasadena, CA 91109
}
\altaffiltext{5}{ 
NASA's Goddard Space Flight Center,
Code 681, 
Greenbelt, MD 20771 
}
\altaffiltext{6}{ 
Lick Observatory,
University of California at Santa Cruz,
1156 High Street,
Santa Cruz, CA 95064
}
\altaffiltext{7}{
Bartko Science \& Technology,
14520 Akron Street, 
Brighton, CO 80602
}
\altaffiltext{8}{
Instituto de Astrof\'{\i}sica de Andaluc\'{\i}a (CSIC),
Camino Bajo de Hu\'{e}tor, 24,
Granada 18008, Spain
}
\altaffiltext{9}{
Department of Physics and Astronomy,
Washington State University, 
Pullman, WA 99164
}
\altaffiltext{10}{
School of Physics and Astronomy, 
Tel Aviv University, 
Tel Aviv 69978, Israel
}
\altaffiltext{11}{
Space Telescope Science Institute,
3700 San Martin Drive, 
Baltimore, MD 21218
}
\altaffiltext{12}{
Metajiva,
12320 Scenic Drive,
Edmonds, WA 98026
}
\altaffiltext{13}{
Conceptual Analytics LLC,
8209 Woburn Abbey Road,
Glenn Dale, MD 20769
}
\altaffiltext{14}{
Royal Observatory Edinburgh,
Blackford Hill,
Edinburgh EH9~3HJ, UK
}
\altaffiltext{15}{
Leiden Observatory,
Postbus 9513, 
2300 RA Leiden, Netherlands
}
\altaffiltext{16}{
Institute of Space and Astronautical Science, 
Japan Aerospace Exploration Agency, 
3-1-1 Yoshinodai, 
Sagamihara, Kanagawa 229-8510, Japan
}
\altaffiltext{17}{
Department of Astronomy and Astrophysics,
The Pennsylvania State University,
525 Davey Lab, 
University Park, PA 16802
}
\altaffiltext{18}{
Departmento de Astronom\'{\i}a y Astrof\'{\i}sica,
Pontificia Universidad Cat\'{o}lica de Chile,
Casilla 306, 
Santiago 22, Chile}
\altaffiltext{19}{
Steward Observatory,
University of Arizona,
Tucson, AZ 85721
}
\altaffiltext{20}{
European Southern Observatory,
Karl-Schwarzschild-Strasse 2, 
D-85748 Garching, Germany
}
\altaffiltext{21}{
W.~M.\ Keck Observatory, 
65-1120 Mamalahoa Hwy, 
Kamuela, HI 96743
}

\begin{abstract} 
We present F435W ($B$), F606W (Broad $V$), and F814W (Broad $I$) coronagraphic images of the debris disk around $\beta$~Pictoris obtained with the 
{\it Hubble Space Telescope's} Advanced Camera for Surveys.  These images provide the most photometrically accurate and morphologically detailed 
views of the disk between 30 and 300~AU from the star ever recorded in scattered light.  We confirm that the previously reported warp in the inner 
disk is a distinct secondary disk inclined by $\sim 5^{\circ}$ from the main disk.  The projected spine of the secondary disk coincides with the 
isophotal inflections, or ``butterfly asymmetry,'' previously seen at large distances from the star.  We also confirm that the opposing extensions 
of the main disk have different position angles, but we find that this ``wing-tilt asymmetry'' is centered on the star rather than offset from it 
as previously reported.  The main disk's northeast extension is linear from 80 to 250~AU, but the southwest extension is distinctly bowed with an 
amplitude of $\sim 1$~AU over the same region.  Both extensions of the secondary disk appear linear, but not collinear, from 80 to 150~AU.  Within 
$\sim 120$~AU of the star, the main disk is $\sim 50$\% thinner than previously reported.  The surface-brightness profiles along the spine of the 
main disk are fitted with four distinct radial power laws between 40 and 250~AU, while those of the secondary disk between 80 and 150~AU are fitted 
with single power laws.  These discrepancies suggest that the two disks have different grain compositions or size distributions.  The F606W/F435W 
and F814W/F435W flux ratios of the composite disk are nonuniform and asymmetric about both projected axes of the disk.  The disk's northwest region 
appears 20--30\% redder than its southeast region, which is inconsistent with the notion that forward scattering from the nearer northwest side of 
the disk should diminish with increasing wavelength.  Within $\sim 120$~AU, the $m_{\rm F435W}$--$m_{\rm F606W}$ and $m_{\rm F435W}$--$m_{\rm F814W}$
colors along the spine of the main disk are $\sim 10$\% and $\sim 20$\% redder, respectively, than those of $\beta$~Pic.  These colors increasingly
redden beyond $\sim 120$~AU, becoming 25\% and 40\% redder, respectively, than the star at 250~AU.  These measurements overrule previous determinations 
that the disk is composed of neutrally scattering grains.  The change in color gradient at $\sim 120$~AU nearly coincides with the prominent inflection 
in the surface-brightness profile at $\sim 115$~AU and the expected water-ice sublimation boundary.  We compare the observed red colors within 
$\sim 120$~AU with the simulated colors of non-icy grains having a radial number density $\propto r^{-3}$ and different compositions, porosities, and 
minimum grain sizes.  The observed colors are consistent with those of compact or moderately porous grains of astronomical silicate and/or graphite with sizes $\gtrsim 0.15$--$0.20~\mu$m, 
but the colors are inconsistent with the blue colors expected from grains with porosities $\gtrsim 90$\%.  The increasingly red colors beyond the 
ice-sublimation zone may indicate the condensation of icy mantles on the refractory grains, or they may reflect an increasing minimum grain size 
caused by the cessation of cometary activity.
\end{abstract}

 
\section{Introduction\label{intro}}
Since the initial discoveries of cool ($\sim 100$~K) dust around nearby stars by the {\it Infrared Astronomical Satellite} \citep{aum85}, 
$\beta$~Pictoris has been the foremost example of a young, main-sequence star with a resolved circumstellar disk of dust.  The disk likely comprises 
the debris from disintegrating bodies in a nascent planetary system rather than primordial dust from the dissipating protostellar nebula \citep{bac93,
art97,lag00,zuc01}.  Spectroscopic evidence of multitudinous star-grazing comets \citep[and references therein]{lag88,beu90,vid94} has motivated models 
of the disk as an admixture of gas and dust from colliding and evaporating comets located within a few tens of AU from $\beta$~Pic \citep{lec96,beu96,
the03}.  The cometary origin of the dust is supported by the detection of broad, $10~\mu$m silicate emission like that observed in the spectra of 
comets Halley, Kohoutek, and others \citep{tel91,kna93,ait93,wei03,oka04}.

Ground-based, coronagraphic images of $\beta$~Pic reveal an asymmetric, flared disk extending at least 1800~AU from the star and viewed nearly 
``edge-on'' \citep{smi84,par87,gol93,kal95,mou97a,lar01}.  High-resolution {\it Hubble Space Telescope (HST)} and adaptive-optics images show that 
the inner part of the disk ($\sim 20$--100~AU from $\beta$~Pic) is warped in a manner consistent with the presence of a secondary disk that is 
inclined by $\sim 4^{\circ}$ from the main disk and perhaps sustained by a massive planet in a similarly inclined, eccentric orbit \citep{bur95,
mou97b,hea00,aug01}.  {\it HST} and ground-based images also reveal concentrations of dust along the northeast extension of the disk about 
500--800~AU from the star that have been interpreted as an asymmetric system of rings formed, along with other asymmetries in the disk, after a 
close encounter with a passing star \citep{kal00,kal01,lar01}.  Spatially resolved mid-infrared images show an asymmetric inner disk having 
depleted dust within 40~AU of $\beta$~Pic \citep{lag94,pan97} and oblique clumps of emission 20--80~AU from the star \citep{wah03,wei03,tel05}.
These features suggest the presence of noncoplanar dust rings whose locations conform to the mean-motion resonances of a putative planetary system.

Constraints on the sizes of the dust grains observed in scattered light have been based upon multiband ({\it BVRI}) imaging studies of 
the disk in both unpolarized \citep{par87,lec93} and polarized \citep{gle91,wol95} light.  The unpolarized images indicate that the 
disk is colorless (within uncertainties of 20--30\%) at distances 100--300~AU from $\beta$~Pic, though its $B$-band brightness may be 
suppressed at 50~AU from the star.\footnote{Throughout this paper, we compute the projected dimensions of the disk in astronomical units 
(AU) using the trigonometric parallax of $\pi = 0\farcs05187~\pm~0\farcs00051$ (or $1/\pi = 19.28 \pm 0.19$~pc) reported for
$\beta$~Pic by \citet{cri97} based on astrometric measurements conducted with the {\it Hipparcos} satellite.  Consequently, the projected distances reported
in this paper may differ from those appearing in papers published before 1997, which were based on an erroneous distance of 16.4~pc to the 
star.}  This neutral scattering by the disk has been customarily viewed as evidence that the dust grains are much larger than the wavelengths
of the scattered light ($\gg 1~\mu$m).  However, \citet{chi91} noted that the $B$-, $V$-, and $I$-band scattering efficiencies of silicate 
spheres were similar for grains with radii of 0.2--$0.3~\mu$m.  Attempts to reconcile the neutral colors of the disk with the 10--25\% 
polarization of scattered light from the disk have been problematic.  \citet{vos99} and \citet{kri00} found that, although the polarization 
alone is best fitted with a grain size distribution with a lower limit of a few microns, the observed neutral colors can only be replicated 
by adding submicron-sized grains and lowering either the refractive index of the grains or the proportion of the smallest grains.  Given the 
adjustments needed to match the polarization models with the highly uncertain disk colors, a more precise multicolor imaging study
of the $\beta$~Pic disk is warranted.

In this paper, we present multiband coronagraphic images of $\beta$~Pic's circumstellar disk obtained with {\it HST's} Advanced Camera for 
Surveys (ACS) \citep{for03,gon05}.  These images reveal the disk between 30 and 300~AU from the star with unprecedented spatial resolution, 
scattered-light suppression, and photometric precision.  These qualities permit the measurement of the disk's optical colors with 3--10 times 
better precision than previously reported from ground-based observations.  By deconvolving the instrumental point-spread function (PSF) from 
each image, we accurately determine the brightnesses, morphologies, and asymmetries of the two disk components associated with the warp in the
inner disk.  Our fully-processed images and results will likely serve as the empirical standards for subsequent scattered-light models of the 
inner disk until the next generation of space-based coronagraphic imagers is deployed.

\section{Observations and Data Processing\label{observations}}

\subsection{ACS Imaging Strategy and Reduction\label{acsimaging}}

Multiband, coronagraphic images of the A5{\footnotesize V} star $\beta$~Pic were recorded on UT~2003 October~1 using the High 
Resolution Channel (HRC) of ACS \citep{for03,gon05}.
The HRC features a $1024 \times 1024$-pixel CCD detector whose pixels subtend an area of 0\farcs028~$\times$~0\farcs025, providing a 
$\sim 29'' \times 26''$ field of view (FOV).  Beta Pic was acquired in the standard ``peak-up'' mode with the coronagraph assembly 
deployed in the focal plane of the aberrated beam.  The star was then positioned behind the small (0\farcs9 radius) occulting spot 
located approximately at the center of the FOV.  {\it HST} was oriented so that the disk's midplane appeared approximately perpendicular
to the $5''$ occulting finger and the large (1\farcs5 radius) occulting spot that also lie in the FOV.  Short, medium, and long exposures
were recorded through the F435W ($B$), F606W (Broad $V$), and F814W (Broad $I$) filters over three consecutive {\it HST} orbits.  All 
images were digitized using the default analog-to-digital conversion of $2~e^-$~DN$^{-1}$.   This sequence of exposures was promptly 
repeated after rolling {\it HST} about the line of sight by $\sim 10^{\circ}$.  This offset changed the orientation of the disk in the 
FOV by $\sim 10^{\circ}$ and facilitated the discrimination of features associated with the disk from those intrinsic to the coronagraphic 
PSF.  Immediately before the exposures of $\beta$~Pic, coronagraphic images of the A7{\footnotesize IV} star $\alpha$~Pictoris were 
recorded through the same filters to provide reference images of a star having colors similar to those of $\beta$~Pic but no known 
circumstellar dust.  A log of all HRC exposures is given in Table~\ref{explog}.

\begin{figure*}[t]
  \epsscale{1.00}\plotone{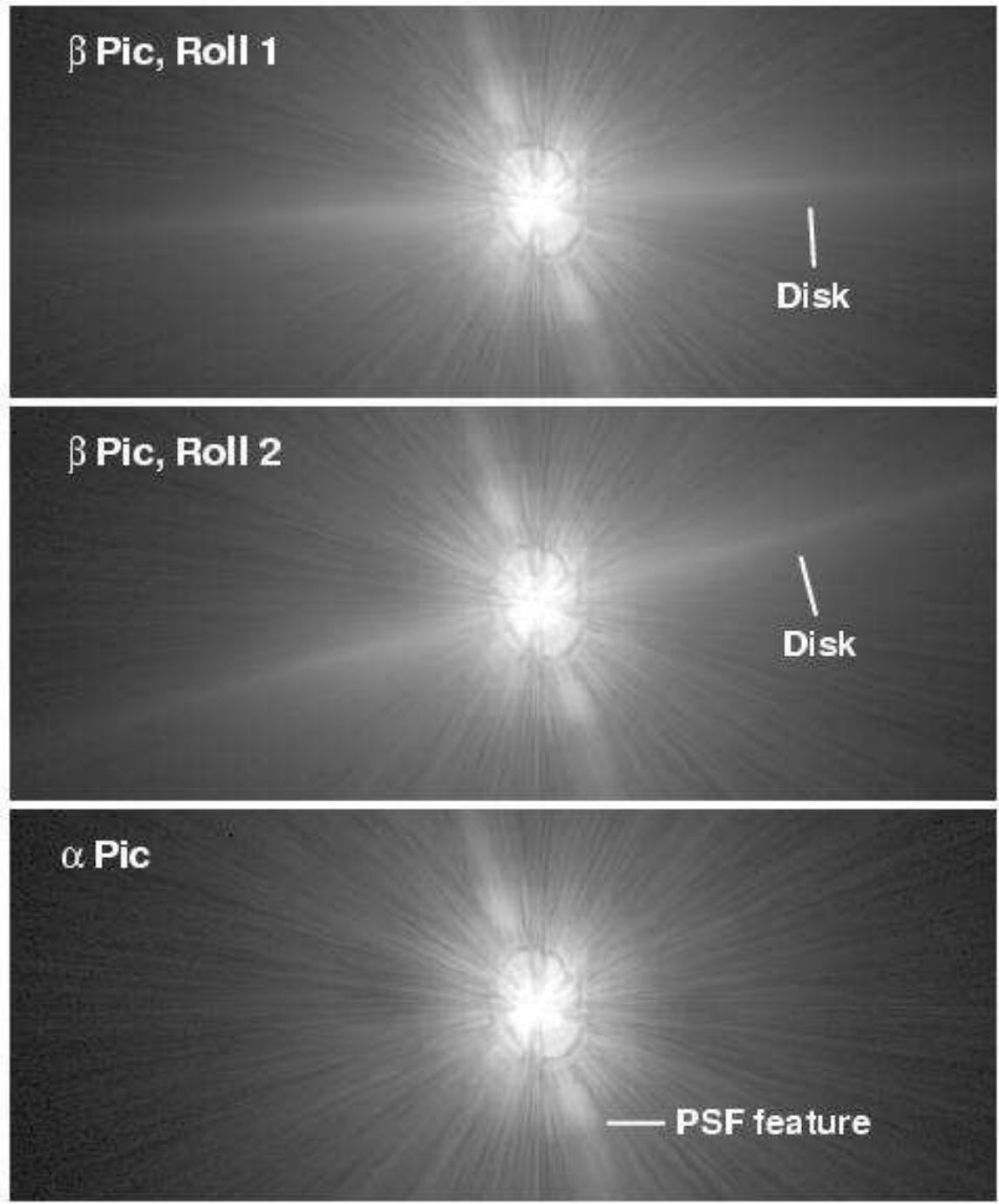}
   \caption{
	$29'' \times 10''$ sections of the F606W (Broad $V$) images of $\beta$~Pic {\it (top and middle panels)} 
	and the reference star $\alpha$~Pic {\it (bottom panel)} obtained with the ACS HRC coronagraph.  The
	images are displayed with logarithmic scaling, but without correction of geometric distortion.  The dust 
	disk around $\beta$~Pic, which is viewed nearly edge-on, is evident without subtraction of the stellar 
	point-spread function.  The apparent position angles of the disk in the top and middle panels change in 
	accordance with the {\it HST} roll offset of $9^{\circ}\!{.}7$.  The linear feature seen in all three 
	panels is a component of the coronagraphic PSF whose origin is presently unknown.  It is collinear with 
	the HRC's occulting finger, which lies beyond the field of view of each panel.
	}
  \label{abpic}
\end{figure*}

The initial stages of image reduction (i.e., subtraction of bias and dark frames and division by a noncoronagraphic flat field) were 
performed by the ACS image calibration pipeline at the Space Telescope Science Institute (STScI) \citep{pav05}.   To correct the
vignetting caused by the occulting spots, we divided the images by normalized ``spot flats'' that were appropriately registered to 
the approximate locations of the migratory occulting spots on the date of our observations \citep{kri04}.  We then averaged the
constituent images of each set of exposures listed in Table~\ref{explog} after interpolating over static bad pixels and eliminating 
transient bad pixels with a conventional $3\sigma$ rejection algorithm.  We then normalized the averaged images to unit exposure time 
and replaced saturated pixels in the long-exposure images with unsaturated pixels at corresponding locations in the shorter-exposure 
images.  Throughout this process, we tracked the uncertainties associated with each image pixel.  In this manner, we created 
cosmetically clean, high-contrast images and meaningful error maps for each combination of star, filter, and roll angle.  
Figure~\ref{abpic} shows $29'' \times 10''$ sections of the reduced F606W images of $\beta$~Pic and $\alpha$~Pic obtained at each 
roll angle.

\subsection{Subtraction of the Coronagraphic PSF\label{psfsub}}

To distinguish the brightness and morphology of the disk from the diffracted and scattered light of $\beta$~Pic, the occulted star's PSF 
must be removed from each image.  By observing $\alpha$~Pic and $\beta$~Pic in consecutive {\it HST} orbits, we limited the differences 
between the coronagraphic PSFs of the two stars that would otherwise be caused by inconsistent redeployment of the coronagraph assembly, 
gradual migration of the occulting spot, or changes in {\it HST's} thermally driven focus cycles \citep{kri02}.  We measured the positions
of the stars behind the occulting spot using the central peaks of the reduced coronagraphic PSFs (Figure~\ref{abpic}) that result from the
reimaging of incompletely occulted, spherically aberrated starlight by ACS's corrective optics \citep{krist00}.  The positions of $\beta$~Pic
and $\alpha$~Pic differed by $\sim 0.8$~pixel ($\sim 0$\farcs02).  This offset causes differences between the coronagraphic PSFs that are 
large in the immediate vicinity of the occulting spot, but the residual light at larger field angles ($\gtrsim 5''$ from the star) after 
PSF subtraction is $\sim 10^{3.5}$ times fainter than the disk's midplane at those field angles \citep{krist00}.

Optimal subtraction of the coronagraphic PSF requires accurate normalization and registration of the filter images of the reference 
star $\alpha$~Pic with the corresponding images of $\beta$~Pic.  Because direct images of the two stars were not obtained, we
estimated the brightnesses of each star in each ACS bandpass using the {\it HST} synthetic photometry package, Synphot, which 
has been developed and distributed by STScI \citep{bus98}.  In doing so, we used the optical spectra of the A5{\footnotesize V} 
stars $\theta^1$~Serpentis and Praesepe~154 \citep{gun83} to approximate the spectrum of $\beta$~Pic.  Likewise, we approximated
the spectrum of $\alpha$~Pic with that of the A5{\footnotesize IV} star HD~165475B.  These proxies yielded synthetic 
Johnson--Cousins photometry that closely match the Cousins {\it BVRI} measurements of $\alpha$~Pic and $\beta$~Pic reported by 
\citet{bes90}.  Assuming $V$ magnitudes of 3.27 and 3.86 for $\alpha$~Pic and $\beta$~Pic, respectively, we computed synthetic
flux ratios, $F_{\alpha}/F_{\beta}$, of 1.65, 1.75, and 1.90 for F435W, F606W, and F814W, respectively.  We then divided the
images of $\alpha$~Pic by these ratios to bring the integrated brightnesses of the reference PSFs into conformity with those
of $\beta$~Pic. 

We aligned the normalized images of $\alpha$~Pic with the corresponding images of $\beta$~Pic using an interactive routine that 
permits orthogonal shifts of an image with subpixel resolution and cubic convolution interpolation.  The shift intervals and 
normalization factors (i.e., $F_{\alpha}/F_{\beta}$) were progressively refined throughout the iterative process.  We assessed
the quality of the normalization and registration by visually inspecting the difference image created after each shift or 
normalization adjustment.  Convergence was reached when the subtraction residuals were visibly minimized and refinements of the
shift interval or normalization factor had inconsequential effects.  Based on these qualitative assessments, we estimate that 
the uncertainty of the registration along each axis is 0.125~pixel and the uncertainty of $F_{\alpha}/F_{\beta}$ in each 
bandpass is 2\%.

After subtracting the coronagraphic PSFs from each image, we transformed the images to correct the pronounced geometric distortion
in the HRC image plane.  In doing so, we used the coefficients of the biquartic-polynomial distortion map provided by STScI
\citep{meu02} and cubic convolution interpolation to conserve the imaged flux.  We then combined the images obtained at each 
{\it HST} roll angle by rotating the images of the second group clockwise by $9^{\circ}\!{.}7$ (Table~\ref{explog}), aligning 
the respective pairs of images according to the previously measured stellar centroids, and averaging the image pairs after 
rejecting pixels that exceeded their local $3\sigma$ values.  Again, we tracked the uncertainties associated with each stage 
of image processing to maintain a meaningful map of random pixel errors.  

We combined in quadrature the final random-error maps with estimates of the systematic errors caused by uncertainties in the 
normalization and registration of the reference PSFs.  Other systematic errors from cyclic changes of {\it HST's} focus and 
differences between the field positions and broadband colors of $\alpha$~Pic and $\beta$~Pic are negligible compared with the 
surface brightness of $\beta$~Pic's disk over most of the HRC's FOV \citep{krist00}.  Our systematic-error maps represent
the convolved differences between the optimal PSF-subtracted image of $\beta$~Pic and three nonoptimal ones generated by purposefully 
misaligning (along each axis) or misscaling the images of $\alpha$~Pic by amounts equal to our estimated uncertainties in PSF 
registration and $F_{\alpha}/F_{\beta}$.  The total systematic errors are 1--5 times larger than the random errors within $\sim 3''$
of $\beta$~Pic, but they diminish to 10--25\% of the random errors beyond $\sim 6''$ of the star.  We refer to the combined maps 
of random and systematic errors as total-error maps.

\begin{figure*}[t]
   \epsscale{1.00}\plotone{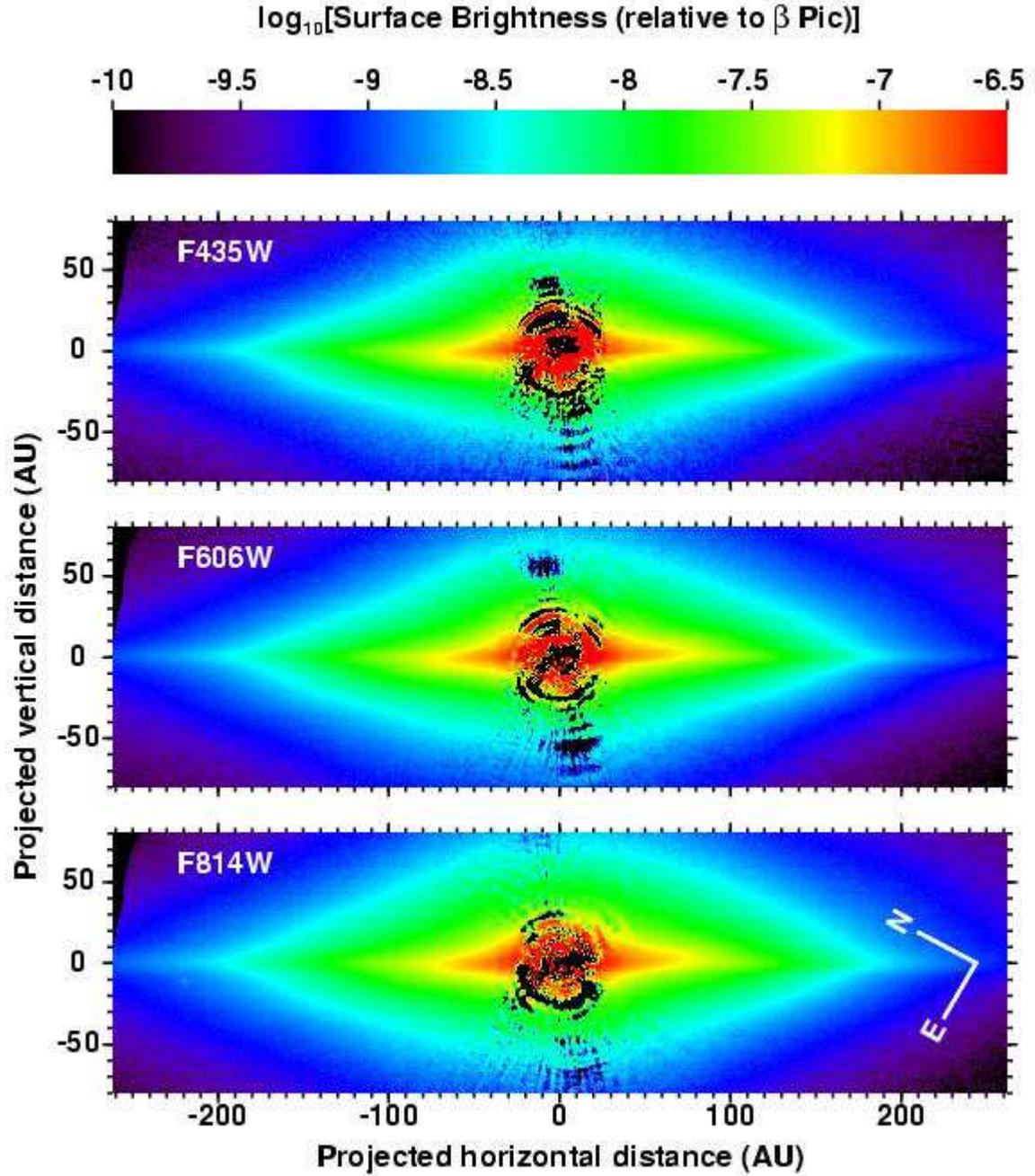}
   \caption{\footnotesize
	Multiband HRC images of the disk around $\beta$~Pic after subtraction of the stellar PSF, but before deconvolution
	of the ``off-spot'' PSF.  The panels show 27\farcs2$~\times$~8\farcs3 sections of the F435W ($B$), F606W (Broad $V$),
	and F814W (Broad $I$) images.  The image sections have been rotated so that the northeast extension of the disk appears
	to the left of each panel and the midplane of the outer disk (radius $\gtrsim 100$~AU) is horizontal.  Each color-coded 
	panel shows the logarithm of the disk's surface brightness relative to the star's brightness in that bandpass.  The 
	irregular, blackened regions near the center of each panel reflect imperfect PSF subtraction around the occulting spot, 
	especially along the direction of the occulting finger, which is beyond the exhibited FOV.  
	}
  \label{bvidisk}
\end{figure*}

Figure~\ref{bvidisk} shows the reduced and PSF-subtracted images of the disk in each ACS bandpass.  Each image has been rotated
so that the northeast extension of the disk is displayed horizontally to the left of each panel.  The images have been divided
by the brightness of $\beta$~Pic in each bandpass derived from Synphot.\footnote{All calibrated surface brightnesses and colors 
presented in this paper are based upon the following Vega-based apparent magnitudes for $\beta$~Pic obtained from 
Synphot: $m_{\rm F435W} = 4.05$, $m_{\rm F606W} = 3.81$, and $m_{\rm F814W} = 3.68$.  The systematic zero-point errors are 
$< 2$\% \citep{sir05}, and the estimated errors from imperfectly matched reference spectra are $\sim 1$--2\%.}  The alternating 
light and dark bands near the occulting spot reflect imperfect PSF subtraction caused by the slightly mismatched colors and 
centroids of $\alpha$~Pic and $\beta$~Pic.  The bands perpendicular to the disk have amplitudes that are $\sim 50$--100\% of the 
midplane surface brightnesses at similar distances from $\beta$~Pic.  These residuals preclude accurate photometry of the disk 
within 1\farcs5 ($\sim 30$~AU) of the star and anywhere along the direction of the occulting finger.  Along the midplane of the 
disk, the photometric uncertainties due to PSF subtraction are $\sim 5$--10\% at a radius of $r = 30$~AU and less than 1\% for 
$r > 60$~AU.

\begin{figure*}[t]
   \epsscale{1.00}\plotone{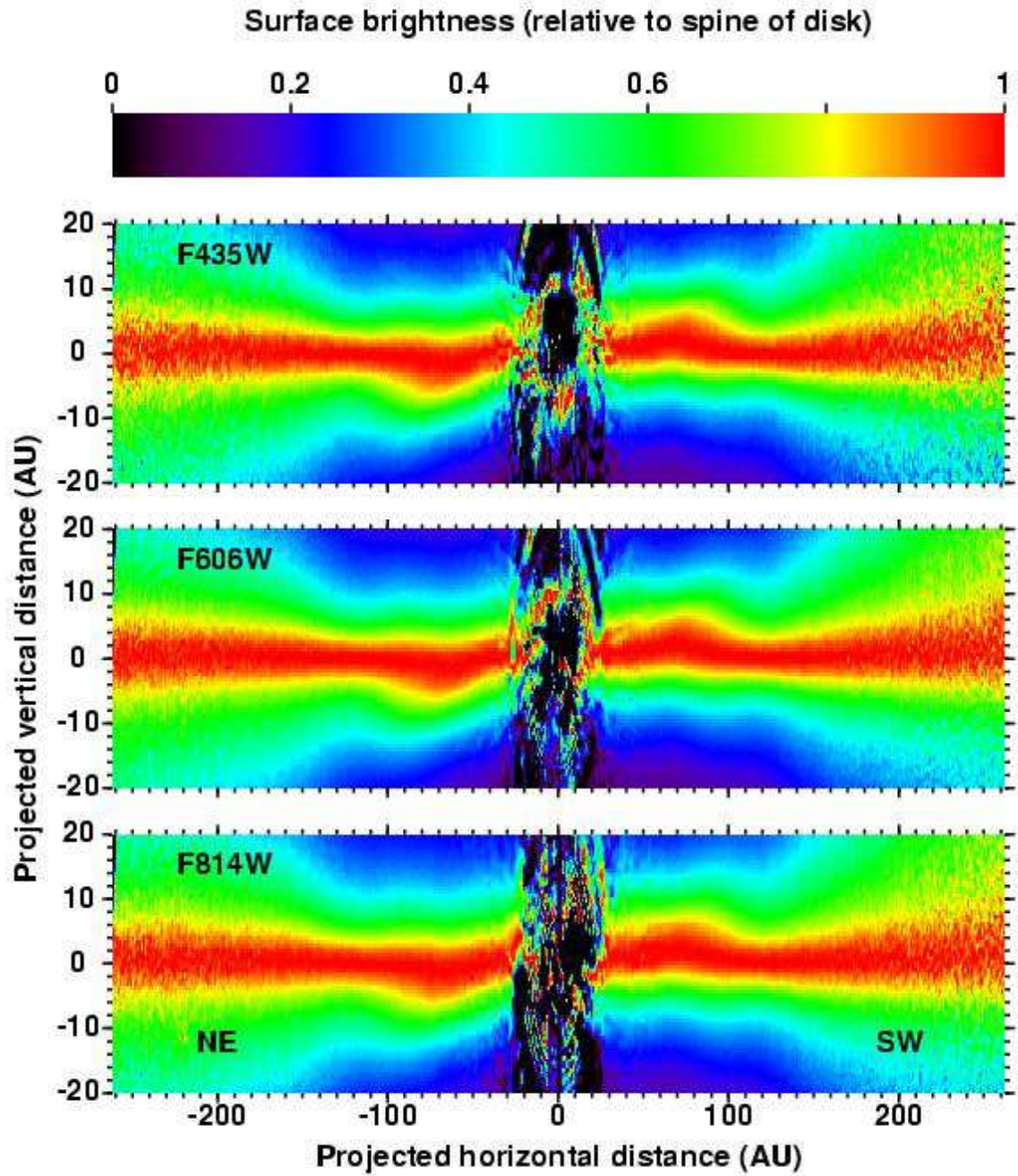}
   \caption{
	Same as Figure~\ref{bvidisk} with the vertical scale expanded by a factor of four.  The color-coded images 
	represent the surface brightnesses of the disk relative to those measured along the spine of the disk.  (See \S\ref{psfsub}
	for details.)  The expanded vertical scale exaggerates the warp in the inner region of the disk first reported 
	by \citet{bur95}.  The red dot seen to the left of the ``NE'' label in the bottom panel is a very red background
	source located 11\farcs5 from $\beta$~Pic at a position angle of $32.0^{\circ}$.
	}
  \label{bvispine}
\end{figure*}

Figure~\ref{bvispine} shows alternate views of the disk in which the vertical scale is expanded by a factor of four over that 
presented in Figure~\ref{bvidisk} and the vertical dimension of the disk's surface brightness is normalized by the brightness measured
along the ``spine'' of the disk.  (The spine comprises the vertical locations of the maximum disk brightness measured along the 
horizontal axis of each image, after smoothing with a $3 \times 3$ pixel boxcar.)  The expanded vertical scale exaggerates the warp
in the inner disk first observed in images taken with {\it HST's} Wide Field Planetary Camera~2 (WFPC2) by \citet{bur95}.  The 
multiband images shown in Figures~\ref{bvidisk} and \ref{bvispine} may be directly compared with the unfiltered optical image of 
the disk obtained with the Space Telescope Imaging Spectrograph (STIS) coronagraph \citep{gra03} and shown in Figure~8 of \citet{hea00}.  

\subsection{Deconvolution of the ``Off-Spot'' PSF\label{psfdecon}}

Accurate assessment of the chromatic dependencies of the disk's color and morphology requires the deconvolution of the unocculted
instrumental PSF from each HRC filter image.  This deconvolution of the ``off-spot'' PSF is especially important for the F814W 
images, because very red photons ($\lambda \gtrsim 0.7~\mu$m) passing through the HRC's CCD detector are scattered diffusely from 
the CCD substrate into a large halo that contributes significantly to the wide-angle component of the PSF \citep{sir05}.  
Unfortunately, no collection of empirical off-spot reference PSFs  exists yet for the HRC coronagraph.  Consequently, we can 
deconvolve the off-spot PSFs only approximately by using synthetic PSFs generated by the Tiny Tim software package distributed by 
STScI \citep{kh04}.  Tiny Tim employs a simplistic model of the red halo that does not consider its known asymmetries \citep{kri05b},
but this model is sufficient for assessing the general impact of the red halo on our images of $\beta$~Pic's disk.

We generated model off-spot PSFs using the optical prescriptions, filter transmission curves, and sample A5{\footnotesize V} source 
spectrum incorporated in TinyTim.  For simplicity, we approximated the weakly field-dependent PSF in each bandpass with single model
PSFs characteristic of the center of the FOV.  The model PSFs extended to an angular radius of $10''$.  We corrected the geometrically 
distorted model PSFs in the manner described in \S\ref{psfsub}, and then deconvolved them from the PSF-subtracted images of $\beta$~Pic
obtained at each roll angle.  In doing so, we applied the Lucy--Richardson deconvolution algorithm \citep{ric72,luc74} to each image 
outside a circular region of radius 1\farcs5 centered on the subtracted star.  (The amplitudes and spatial frequencies of the 
PSF-subtraction residuals within this region were too large to yield credible deconvolved data.)  The imaged FOV lacked any 
bright point-sources by which we could judge convergence of the deconvolution, so we terminated the computation after 50 iterations.
Examination of intermediate stages of the process showed no perceptible change in the deconvolved images after $\sim 45$ iterations.

Figures~\ref{bvidisklucy} and \ref{bvispinelucy} are the deconvolved counterparts of Figures~\ref{bvidisk} and \ref{bvispine}, 
respectively.  The compensating effect of the deconvolution is especially evident when comparing the morphologies of the disk in 
Figures~\ref{bvispine} and \ref{bvispinelucy}.  The color-coded isophotes of the disk in each bandpass are much more similar in the 
deconvolved images than in the convolved images.  The amplified, correlated noise in the deconvolved images is characteristic
of the Lucy--Richardson algorithm when applied to faint, extended sources.  It is a consequence of the algorithm's requirements of a
low background signal, nonzero pixel values, and flux conservation on both local and global scales.  These requirements also account
for the disappearance of the negative PSF-subtraction residuals near the occulting spot.  The requirement of local flux conservation
ensures reliable photometry in regions where the ratio of signal to noise (S/N) is large and where PSF-subtraction residuals are 
small, but it makes photometric measurements elsewhere less accurate and their uncertainties nonanalytic.

\section{Image Analysis\label{imanal}}

The processed ACS/HRC images shown in Figures~\ref{bvidisk} and \ref{bvispine} are the finest multiband, scattered-light images of 
$\beta$~Pic's inner disk obtained to date.\footnote{Fully processed images and error maps in FITS format can be obtained via the 
World Wide Web at http://acs.pha.jhu.edu/$\sim$dag/betapic.}  Earlier ground-based, scattered-light images show the 
usual effects of coarse spatial resolution and PSF instability caused by variable atmospheric and local conditions 
\citep{smi84,par87,gol93,lec93,kal95,mou97a, mou97b}.
The unpublished {\it BVRI} WFPC2 images of the disk described by \citet{bur95} have comparatively low S/N ratios because WFPC2 lacks 
a coronagraphic mode and directly imaged starlight scatters irregularly along the surface of its CCD detectors.  These conditions 
forced short exposure times to avoid excessive detector saturation and created irreproducible artifacts in the PSF-subtracted 
images.  Moreover, the construction of a WFPC2 reference PSF from images of $\beta$~Pic obtained at several roll angles allowed 
possible contamination of the reference PSF by the innermost region of the disk.  The unfiltered STIS images 
of the disk \citep{hea00} compare favorably with our HRC images, notwithstanding their lack of chromatic information and partial
pupil apodization \citep{gra03}.  Both sets of images show the same region of the disk, though STIS's narrow occulting wedge 
permitted imaging of the disk's midplane about 0\farcs4 (8~AU) closer to the star.  The HRC's circular occulting spot and Lyot stop
exposed the regions around the projected minor axis of the disk that were obscured in the STIS images by diffraction spikes and the
shadow of the occulting bar.  The HRC images have twice the spatial resolution of the STIS images, and they exhibit better S/N 
ratios in the regions of the disk between 150 and 250~AU from the star.

\subsection{Disk Morphology\label{diskmorph}}

Figure~\ref{diskcon} shows isophotal maps of our F606W images of the disk before and after deconvolution of the off-spot PSF.  These maps 
are qualitatively similar to those generated from our F435W and F814W images.  They confirm the morphology and asymmetries of the disk 
reported previously by several groups and described in detail by \citet{kal95} and \citet{hea00}.  The brightness asymmetry between the 
disk's opposing extensions is particularly evident along the spine beyond $\sim 100$~AU of $\beta$~Pic.  The asymmetric curvature of the 
isophotes on opposite sides of the spine and the inversion of this asymmetry across the projected minor axis of the disk are also apparent.  
\citet{kal95} referred to this diametrical antisymmetry as the ``butterfly asymmetry.''

Our images provide the first credible look at the surface brightness along the disk's projected minor axis.  
The convex isophotes in this region indicate that the pinched appearances of the disk in the images of \citet{smi84}, \citet{gol93}, 
\citet{lec93}, \citet{mou97a,mou97b}, and, to a lesser degree, \citet{kal95} are artifacts of oversubtraction of the reference PSF 
and/or self-subtraction of the disk along its minor axis.  Other scattered-light images yielded no information in this region because 
of obscuration by coronagraphic masks \citep{par87,gle91,hea00}.  Our images are tinged by residuals from the imperfect 
subtraction of the linear PSF feature seen in Figure~\ref{abpic}, but these residuals do not affect the overall contours of the isophotes.
The isophotes are more widely spaced along the northwestern semiminor axis than along the southeastern semiminor axis.  Such brightness 
asymmetry in optically thin disks is often attributed to the enhanced forward-scattering efficiency of the dust, which implies that the 
side of the disk closer to Earth is inclined slightly northwest from the line of sight to the star.  This deduction is consistent with 
the inclination determined by \citet{kal95} from single-scattering models of the surface brightness along the projected major axis of the disk.  

\subsubsection{The Main and Secondary Disks\label{mainsec}}

\citet{mou97b} viewed the inner warp as a deformed and thickened region of the disk, perhaps caused and sustained by a planet in an inclined, 
eccentric orbit within 20~AU of $\beta$~Pic.  \citet{hea00} subsequently interpreted the warp as a blend of two separate disk components,
the lesser of which is inclined from the main component by $4^{\circ}\!{.}6$.  The latter interpretation is supported by Figure~\ref{imratio},
which shows the ratios of our HRC images after and before deconvolution of the off-spot PSF.  The division of the PSF-deconvolved filter 
images (Figure~\ref{bvidisklucy}) by their corresponding convolved images (Figure~\ref{bvidisk}) accentuates the small or narrow features 
in the disk that are most affected by blurring from the instrumental PSF.  Thus the sharply-peaked midplane of the outer disk appears 
prominently in Figure~\ref{imratio}.  The midplane of the inner warp is not as well defined as that of the outer disk, but the apparent 
separation of the midplanes beyond $\sim 80$~AU from the star indicates that the warp is a distinct secondary disk inclined from, and 
perhaps originating within, the main disk.

Mimicking \citet{hea00}, we determined the contributions of the main and secondary disks to the composite vertical scattered-light
profile (i.e., the scattered-light distribution perpendicular to the midplane of the disk) by fitting two similar, vertically 
symmetric profiles to the composite profile as a function of distance from the star.  In doing so, we assigned a shape to the 
individual profiles by qualitatively assessing various analytic functions at a few locations along the disk in our PSF-deconvolved
images.   We found that a ``hybrid-Lorentzian'' function -- two Lorentzians of different widths whose top and bottom parts, 
respectively, are smoothly joined at an arbitrary distance from their common centers -- satisfactorily matched each individual 
profile.  (We ascribe no physical significance to this hybrid-Lorentzian function, nor do we claim that it uniquely or optimally 
characterizes each profile.)  We used a non-linear least-squares fitting algorithm based on the Levenberg--Marquardt technique 
\citep{pre92} to determine the positions, amplitudes, widths, and top-to-bottom transition zones of the main and secondary 
hybrid-Lorentzian profiles at distances of 30--250~AU from $\beta$~Pic.  The resulting fits of the composite vertical profiles 
were best in the regions 80--130 AU from the star, where the two disk components are sufficiently bright and separated to allow 
unambiguous discrimination of both components.  Figure~\ref{vertprof} shows the best fits to the composite vertical profiles of 
both extensions of the disk observed in our F606W image at distances of 100~AU from the star.

Figure~\ref{traceprof} shows the relative positions of the fitted components of the composite vertical profile as a function 
of distance from $\beta$~Pic.  For each filter image and disk extension, we performed linear least-squares fits to the traces 
of the main and secondary disks within the regions where they are accurately identified (80--250~AU for the main disk and 
approximately 80--130~AU for the secondary disk).  These fits are overplotted in Figure~\ref{traceprof}.  The relative position 
angles of the main disk's northeast and southwest extensions (measured counterclockwise from the horizontal axis) are about 
$-0^{\circ}\!{.}4$ and $0^{\circ}\!{.}5$, respectively.  The difference between these position angles is similar to the 
``wing-tilt asymmetry'' noted by \citet{kal95} at distances of 100--450~AU from $\beta$~Pic.  However, our fitted lines appear 
to intersect at a point almost coincident with the star, rather than at a point in the southwest extension, as reported by 
\citeauthor{kal95}.  Expanding the vertical scale of our traces of the main disk in the F606W image (Figure~\ref{maintrace}) 
shows that the spine of the northeast extension conforms to the linear fit down to $\sim 80$~AU, at which point the main and 
secondary disks can no longer be credibly distinguished.  However, the spine of the southwest extension exhibits more curved 
than linear behavior over the range of the fit.  This bow in the southwest extension appears in our F435W and F814W images with
equal clarity and therefore does not appear to be an artifact of our decomposition of the vertical scattered-light profile.

Table~\ref{tiltang} lists the position angles of the linear fits along each extension of the secondary disk relative to the 
corresponding fits along the main disk.  The average tilt of the secondary disk's southwest extension matches the 
$4^{\circ}\!{.}6$ derived by \citet{hea00} for both extensions.  However, we find that the average tilt of the northeast 
extension is $\sim 25$\% larger than the values obtained by \citet{hea00} and by us for the southwest extension.  Extrapolating the 
linear fits to the secondary disk outward (Figure~\ref{diskcon}) shows that the convex inflections of the butterfly asymmetry 
are aligned with the secondary disk, which supports the notion that the asymmetry is caused by radiatively expelled dust from 
the secondary disk \citep{aug01}.  On the other hand, extending the linear fits to the secondary disk toward $\beta$~Pic 
(Figure~\ref{traceprof}) shows that its nonparallel extensions would not converge near the star if they extended sufficiently 
inward.  Instead, the northeast extension would intersect the northeast extension of the main disk at a distance of $\sim 30$~AU
from $\beta$~Pic.  Coronagraphic images with a smaller occulting spot and higher pixel resolution are needed to determine if 
this intersection actually occurs.

Although our analysis of the two-component structure of the disk qualitatively supports that of \citet{hea00}, our results
reveal some errors in the presentation and interpretation of their results.  These errors stem from the incorrect portrayal 
of the secondary disk's tilt relative to the northeast and southwest extensions of the main disk in Figures~11--13 of \citet{hea00}.
In these figures, the plotted data -- and consequently the relative slope of the secondary disk -- are inverted about the 
midplane of the main disk.  The orientations of the main and secondary disks should appear as they do in 
Figures~\ref{imratio}--\ref{traceprof} of this paper and in Figures~8 and 10 of \citet{hea00}.  Because of this inversion, 
\citeauthor{hea00} concluded that the disk's southeast side was brighter than its northwest side and, consequently, that the 
southeast side was inclined toward the line of sight.  These conclusions are not supported by our isophotal map of the disk 
(Figure~\ref{diskcon}) or by the single-scattering models of \citet{kal95}.

\subsubsection{Thickness of the Main Disk\label{diskwidth}}

Previous scattered-light studies indicated that the composite disk is uniformly thick within $7''$ (135~AU) of $\beta$~Pic and 
flared beyond that region \citep{smi84,par87,art90,gol93,kal95,mou97a,hea00}.  These characteristics are evident in Figures~\ref{bvispine} 
and \ref{bvispinelucy}.  Previous measurements of the disk's full width at half its midplane (FWHM) brightness vary significantly, 
however, which suggests that the disk's intrinsic projected width is less than or comparable to the resolution of the images from
which the measurements were made.  Our deconvolution of the off-spot PSF from the HRC images -- the first such effort exhibited for 
scattered-light images of $\beta$~Pic's disk -- permits us to assess this resolution dependency and to measure more accurately the 
intrinsic width of the disk as a function of distance from $\beta$~Pic. 

Figure~\ref{diskwidth606} shows the full widths of each extension of the composite disk at half and one tenth of the maximum 
brightness of the midplane at distances 20--250~AU from the star.  These widths are shown in the panels labeled ``FWHM'' and 
``FW0.1M,'' respectively.  The associated pairs of thin solid and dashed curves show the widths of the northeast and southwest 
extensions obtained from the F606W image before and after PSF deconvolution.  The short, thick curves show the measured widths
of the hybrid-Lorentzian curves that best fit the main component of the composite disk in the region 80--150~AU from $\beta$~Pic 
(\S\ref{mainsec}).  The FWHM and FW0.1M plots show progressive reductions in the measured width of the main disk after PSF deconvolution
and decomposition of the main and secondary disks.  The up-to-30\% reduction in the FWHM after deconvolution of the narrow 
off-spot PSF (FWHM $\approx$~0\farcs05) shows that previously reported measurements, particularly those obtained from the ground, 
reflect more the angular resolution of the observations than the intrinsic thickness of the disk.

The deconvolved and decomposed curves show that the FWHM and FW0.1M of both disk extensions are nearly constant ($\sim 11$--13~AU
and $\sim 50$--60~AU, respectively) within $\sim 120$~AU of $\beta$~Pic, and they increase approximately linearly with distance beyond 
$\sim 120$~AU.  This behavior is consistently seen in the F435W, F606W, and F814W images.  The southwest extension appears to thicken 
more rapidly than the northeast extension beyond $\sim 160$~AU, as noted by \citet{kal95}, but the correlated noise injected into 
the deconvolved images by the Lucy--Richardson algorithm (Figure~\ref{bvispinelucy}) precludes accurate quantification of the
FWHMs from those images.  Reverting to the PSF-convolved images, we find that the FWHM and FW0.1M of the southwest extension are
$\sim 10$~AU and $\sim 60$~AU larger, respectively, than the corresponding values of northwest extension 250~AU from the star.
Our finding is consistent with the FWHM measurements of \citet{kal95} when the different angular resolutions of the two sets of 
observations are considered.  \citet{hea00} reported no such thickness asymmetry from their STIS images.  Their measurements
of FWHM at discrete locations 50--200~AU from $\beta$~Pic are consistent with our PSF-convolved measurements of the narrower northeast
extension.  However, their FW0.1M measurements at these locations are 20--60\% smaller than our PSF-convolved measurements at the same
locations in each extension.  These discrepancies may reflect differences between the S/N ratios of the HRC and STIS images at 
scattered-light levels below $\sim 10^{-9}$ of the stellar flux.

\subsubsection{Planetesimal Belts or Rings\label{rings}}

Reexamining earlier WFPC2 and ground-based coronagraphic images of $\beta$~Pic, \citet{kal00} noted several localized 
density enhancements along the midplane of the disk's northeast extension at distances of 500--800~AU from the star.  They 
did not see this structure in the southwest extension, so they interpreted the clumps as a system of eccentric, nested
rings formed in the aftermath of a close encounter between the disk and a passing star.  \citet{wah03} and \citet{wei03}
later reported mid-infrared images of the disk that show a distinct warp within 20~AU of the star that is oppositely tilted
from the secondary disk seen in Figure~\ref{imratio}.  \citet{wah03} also noted clumps of emission in both disk extensions 
within 100~AU of $\beta$~Pic that are arranged in diametrically opposite pairs centered on the star.  They interpreted these 
clumps as a series of noncoplanar rings caused by gravitational interactions of the disk with a planetary system.  \citet{tel05}
did not observe the innermost warp in their mid-infrared images of the disk, but they attributed a bright clump of 12--$18~\mu$m
emission at 52~AU in southwest extension to a concentration of 0.1--$0.2~\mu$m silicate and organic refractory grains heated to 
$\sim 190$~K.  High-spatial-resolution, mid-infrared spectra of the disk recorded at 3~AU intervals along the inner 30~AU of each
extension also show concentrated regions of emission from 0.1--$2~\mu$m silicate grains \citep{oka04}.  The presence of such 
concentrations of small grains in the face of expulsive stellar-radiation pressure suggests that the grains are continuously 
created by collisions of planetesimals confined to rings or belts within the otherwise depleted inner region of the dust disk.

The bright clumps detected by \citet{kal00} in the northeast extension of the disk lie outside the FOV of our HRC images, 
and the innermost warp reported by \citet{wah03} and \citet{wei03} is obscured by the HRC's occulting spot.  However, the 
clumps of mid-infrared emission located 30--100~AU from the star \citep{wah03,tel05} are potentially observable in our HRC 
images.  Figure~\ref{ringcon} is an isophotal map of the inner region of the disk obtained from our PSF-deconvolved F606W image
after smoothing with a $3 \times 3$-pixel boxcar.  The isophotal interval is 0.2~mag~arcsec$^{-2}$, which is similar to the 
interval between isophotes of $18~\mu$m emission from the same region of the disk presented by \citet{wah03} and \citet{tel05}.
The locations of the clumps of $18~\mu$m emission are marked with the letters assigned to them by \citeauthor{wah03} ~Our 
F606W isophotes show no evidence of such clumping in scattered optical light.  Adjusting the isophotal interval and the 
smoothing factor does not alter this conclusion.  Our images do not refute the existence of the clumps, however, because 
the scattered optical light and mid-infrared emission may emanate from different regions of the disk \citep{pan97}.

\subsection{Surface Brightness and Asymmetry\label{surfbright}}

Because $\beta$~Pic's disk is viewed nearly ``edge-on,'' its surface brightness has traditionally been parametrized with
one or more power laws fitted along the midplanes (or spines) of its opposing extensions \citep{smi84,art89,art90,gle91,
gol93,lec93,kal95,mou97a,hea00}.  The number of fitted power laws and their indices have varied considerably, mostly because 
such fits are sensitive to errors in the stellar PSF subtraction.  The innate stability of {\it HST's} instrumental PSFs 
reduce these errors substantially, so we compare our quantitative measurements of the disk's spine (as defined in \S\ref{psfsub}) 
with those obtained from the STIS images of \citet{hea00}.  To do so, we assess the surface brightnesses of the spines of 
both extensions from our ACS images both before and after PSF deconvolution.  We initially ignore the two-component structure
of the disk described in \S\ref{mainsec}, because the two components cannot be distinguished by our profile-fitting algorithm
within 80~AU of the star (\S\ref{mainsec}) and because the main disk dominates the surface brightness beyond 80~AU.  We then 
compare our results for the composite disk with those obtained for its main and secondary components over the region in which 
they are credibly resolved.

\subsubsection{Horizontal Profiles of the Composite Disk\label{compositeprof}}

Figure~\ref{midpfit} shows the logarithmic surface brightness profiles measured along the spines of each extension of the
composite disk before PSF deconvolution.  The accompanying $\pm 1\sigma$ error profiles, which are derived from the 
total-error maps of the images (\S\ref{psfsub}) and include $\sim 3$\% uncertainty in the photometric calibration, 
show that these profiles are very accurate beyond $\sim 1$\farcs5 ($\sim 30$~AU) from the 
star.  The profiles clearly do not indicate a single power-law dependence of surface brightness with distance, as was 
determined in some of the earliest ground-based imaging studies of the disk \citep{smi84,art89,gle91,lec93}.  Instead, the 
logarithmic profiles exhibit distinctly different linear behavior at angular distances of $2''$--3\farcs5, 3\farcs7--5\farcs6, 
6\farcs6--$10''$, and $10''$--13\farcs4, with smooth transitions in slope between these regions.  Table~\ref{powlaw1} lists 
the slopes of the linear least-squares fits to the logarithmic data (i.e., the indices of the radial power laws, $r^{-\alpha}$, 
that best fit the surface-brightness profiles) within these four regions.  The prominent changes in $\alpha$ at $\sim 6''$ 
($\sim 115$~AU) from the star are well documented \citep{art90,gol93,kal95,hea00}, but the more subtle changes at $\sim 3$\farcs6 
($\sim 70$~AU) and $\sim 10''$ ($\sim 195$~AU) are noted here for the first time.  Conversely, the pronounced decrease in the 
disk's $B$-band surface brightness within 7\farcs3 ($\lesssim 140$~AU) reported by \citet{lec93} is not observed in our F435W 
image.

The prominent inflections in the profiles at $\sim 115$~AU are possible evidence for the sublimation of water ice from composite grains 
at dust temperatures greater than 100--150~K \citep{nak88,art90,gol93,pan97,li98}.  Applying their cometary dust model to $\beta$~Pic's disk, 
\citet{li98} noted that the relative compositions of the grains should vary across several regions of the disk, as the inner and outer 
mantles of various ices and organic refractory material successively evaporate from, or condense upon, their silicate cores in accordance
with the local dust temperature and stellar-radiation pressure.  The more subtle changes in $\alpha$ at $\sim 70$~AU and $\sim 195$~AU 
may therefore mark the boundaries over which such changes in composition (and consequently albedo) occur.  On the other hand, the decrease
in $\alpha$ within $\sim 70$~AU may indicate a decrease in the number density of the grains caused by an interior planetary system 
\citep{lag94,roq94,laz94} or, as we discuss in \S\ref{componentprof}, the unresolved superposition of the main and secondary disks. 

The changes in $\alpha$ at angular distances of $\sim 3$\farcs6, $6''$, and $10''$ from $\beta$~Pic are also apparent in the 
midplane surface-brightness profiles extracted from the STIS images of the disk \citep{hea00}.  To facilitate comparison with previous 
studies, \citeauthor{hea00}\ fitted power-law functions to three regions of each extension.  Two of these regions coincide approximately
with our regions~1 and 3 defined in Table~\ref{powlaw1}; the third region bridges our regions~1 and 2.  Our values of $\alpha$ in region~1
($2''$--3\farcs5, or 39--67~AU, from $\beta$~Pic) of both extensions differ by $\lesssim 5$\% from those computed by \citet{hea00} for 
the positions of maximum flux along each extension.  However, our values of $\alpha$ in region~3 (6\farcs6--$10''$, or 127--193~AU, 
from $\beta$~Pic) are $\sim 10$--15\% smaller than those of \citeauthor{hea00} for both extensions, i.e., our power-law fits in this 
region are significantly less steep than those determined from the STIS images.  This discrepancy is puzzling, especially as this outer 
region of the disk is relatively insensitive to moderate errors in the PSF subtraction.  

To investigate the possible effect of the convolved instrumental PSF on our power-law fits, we repeated this analysis on our 
PSF-deconvolved images in each bandpass.  The surface-brightness profiles along the spines of each deconvolved disk extension 
are shown in Figure~\ref{midpfitlucy}.  The corresponding values of $\alpha$ for each region of the deconvolved images are listed 
in Table~\ref{powlaw1}.  These values are significantly different from their PSF-convolved counterparts only in region~4, where
the S/N ratios are low and, consequently, the efficacy of the Lucy--Richardson algorithm is suspect.  In region~3, the average 
difference between the associated convolved and deconvolved indices is $\sim 1$\%, so it is unlikely that differences between the 
HRC and STIS coronagraphic PSFs cause the $\sim 10$--15\% discrepancies in the values of $\alpha$ reported by \citet{hea00} and us 
for this region.

Figure~\ref{midpsblucy} displays the PSF-deconvolved profiles in a manner that facilitates comparison of the two disk extensions in 
the F435W, F606W, and F814W images.  The plots show the effects of the asymmetric values of $\alpha$ on the relative brightnesses of
the two extensions at equal distances from $\beta$~Pic.  The southwest extension is brighter than the northwest extension in the region
$\sim 50$--100~AU from the star, whereas the opposite condition exists for $r \gtrsim 150$~AU.  At $r = 6''$ ($\sim 115$~AU), the 
surface brightnesses of each extension are 14.9, 14.6, and 14.4~mag~arcsec$^{-2}$ in F435W, F606W, and F814W, respectively.  These values
are 0.2--0.3~mag~arcsec$^{-2}$ brighter than the corresponding brightnesses measured from the PSF-convolved images.  The latter 
measurements compare favorably with the ground-based, $R$-band measurements of \citet{gol93} and \citet{kal95} at $r = 6''$, but are 
$\sim 1.5$~mag~arcsec$^{-2}$ brighter than the $R$- and $I$-band measurements reported by \citet{par87} and \citet{smi84}, respectively,
for this fiducial distance.  The outer distance limits of our photometric measurements (260~AU and 300~AU for the northeast and southwest 
extensions, respectively) are set by the HRC's FOV; they do not reflect an intrinsic asymmetry in the physical sizes of the extensions.

Although the values of $\alpha$ within $\sim 110$~AU of $\beta$~Pic (regions~1 and 2) are nearly constant across F435W, F606W, and F814W, 
those between 125 and 200~AU (region~3) decrease nonuniformly with increasing wavelength.  These trends indicate that the optical colors 
of the disk are constant within $\sim 110$~AU of the star, but they redden considerably between 125 and 200~AU.  The color gradient appears
to flatten again from $\sim 200$ to 250~AU (region~4), but the relatively large uncertainties attached to $\alpha$ in this region preclude
a definitive assessment.  We confirm these trends in our subsequent analysis of the disk's color (\S\ref{optcolor}).

\subsubsection{Horizontal Profiles of the Component Disks\label{componentprof}}

The resolution of the inner disk into two components prompts an assessment of the contributions of each component to the horizontal
surface-brightness profiles of the composite disk.  Figure~\ref{mainsecprof} shows the multiband, logarithmic profiles of the main and 
secondary disks,
extracted from the PSF-deconvolved images.  The curves trace the maxima of the hybrid-Lorentzian profiles that best fit the vertical 
surface-brightness profiles of the two disks at 80--150~AU from $\beta$~Pic (\S\ref{mainsec}).  The secondary disk is $\sim 1.5$~mag~arcsec$^{-2}$
fainter than the main disk at 100~AU, as indicated in Figure~\ref{vertprof}.  The profiles of the main disk show the same inflections at 
$\sim 6''$ ($\sim 115$~AU) exhibited by the composite disk, but the secondary disk's profiles appear linear throughout this region.  
Whereas the relative brightnesses of the main disk's extensions invert at $\sim 115$~AU (as noted in \S\ref{compositeprof} for the 
composite disk), the southwest extension of the secondary disk is consistently brighter than its northeast extension from 80 to 150~AU.

We fitted power laws to the profiles of both extensions of the main and secondary disks over the regions displayed in Figure~\ref{mainsecprof},
taking care to avoid the inflection of the main disk at $\sim 115$~AU.  Table~\ref{powlaw2} lists the indices, $\alpha$, of these power 
laws.  The main disk's indices at 80--108~AU are 15-35\% smaller than those of the composite disk in the mostly overlapping region~2
(defined in Table~\ref{powlaw1}), whereas its indices at 127--150~AU are 5--20\% larger than those of the composite disk in region~3.
These indices are nearly constant across the $B$, $V$, and $I$ bands at 80--108~AU, but they progressively decrease with increasing 
wavelength at 127--150~AU.  This behavior mimics that of the composite disk in regions~2 and 3 (\S\ref{compositeprof}), which indicates 
that the main disk's colors progressively redden beyond the inflection at $\sim 115$~AU.  Conversely, the secondary disk's indices 
progressively increase with increasing wavelength within the entire 80--150~AU region of the northeast extension, but they are effectively
constant in the southwest extension.  The former trend suggests that the colors in the northeast extension become bluer with increasing 
distance from $\beta$~Pic.  Altogether, these characteristics indicate that the grain populations and/or distributions in the 
main and secondary disks differ significantly at common projected distances from the star.  Moreover, if unresolved, the secondary disk 
can significantly affect models that use the scattered-light profiles to constrain the spatial distribution and composition of dust in 
the main disk.

\subsection{Optical Colors of the Disk\label{optcolor}}

Previous multiband-imaging studies of $\beta$~Pic's disk showed that the optical colors of the star and disk are indistinguishable within 
the studies' estimated photometric errors \citep{par87,lec93}.  These errors ranged from 20 to 30\%, so actual color differences of a few 
tenths of a magnitude would not have been deemed credible.  The greater photometric precision of our ACS study enables a correspondingly
more precise assessment of the disk's optical colors and scattering properties.  We first examine the ACS colors of the composite disk 
to enable a comparison with the earlier ground-based studies.  We then examine the colors of the main and secondary components derived from
our decomposition of the disk's vertical scattered-light profile (\S\ref{mainsec}).

\subsubsection{Two-Band Flux Ratios of the Composite Disk\label{fluxrat}}

\begin{figure*}[t]
   \epsscale{1.00}\plotone{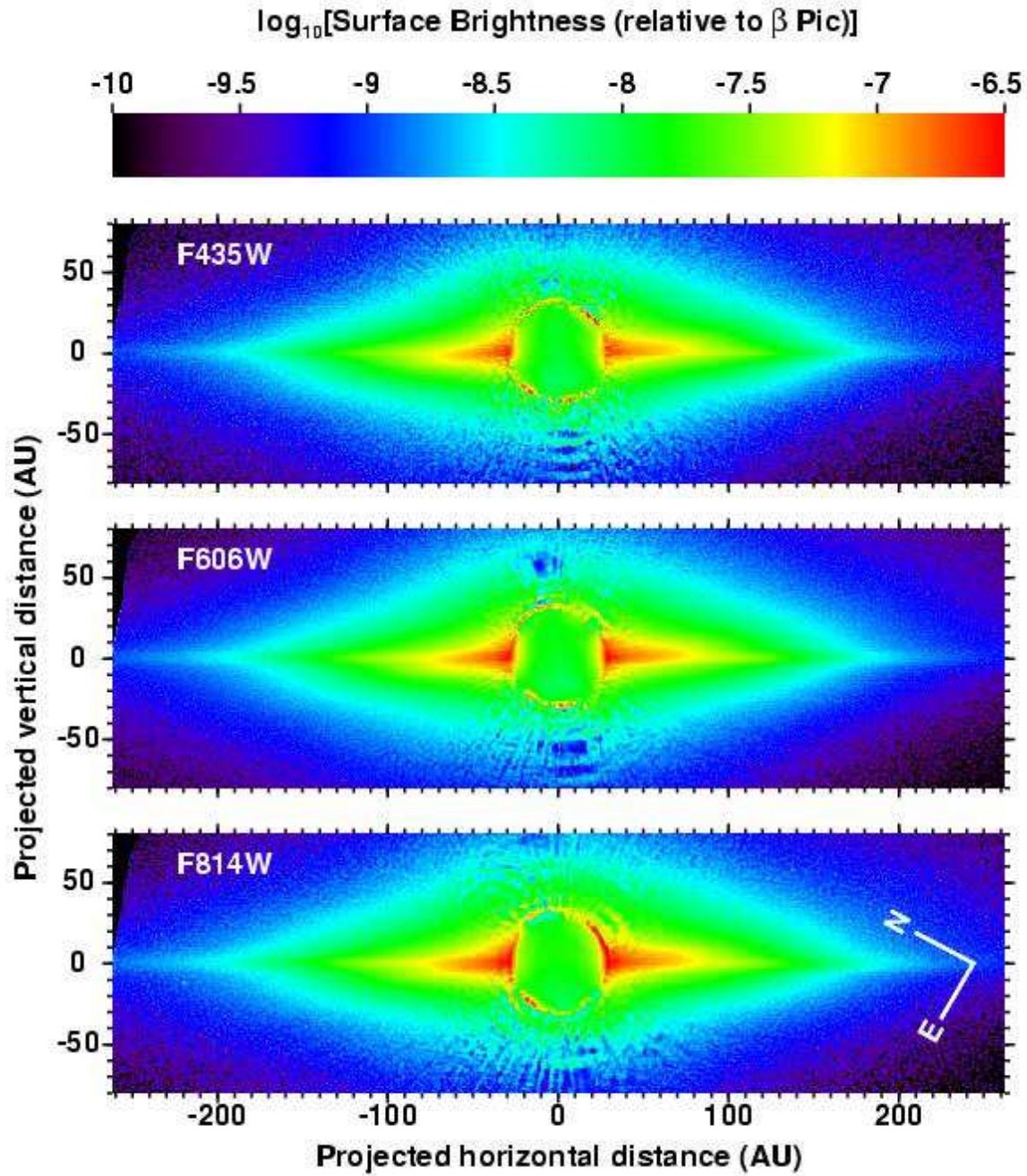}
   \caption{
	Same images shown in Figure~\ref{bvidisk} after Lucy--Richardson deconvolution of the ``off-spot'' PSF.  Pixels 
	lying within a radius of 1\farcs5 ($\sim 30$~AU) of the image center (i.e., the location of the occulted star) have been 
	masked and excluded from the deconvolution.  Because the Lucy--Richardson algorithm forces all pixels to have positive
	values, the deconvolved images exhibit no negative PSF-subtraction residuals and enhanced, correlated noise at faint 
	signal levels.  Consequently, photometry of the disk in these regions of the disk is less reliable than in regions 
	having large S/N ratios and small subtraction residuals.
	}
  \label{bvidisklucy}
\end{figure*}

\begin{figure*}[t]
   \epsscale{1.00}\plotone{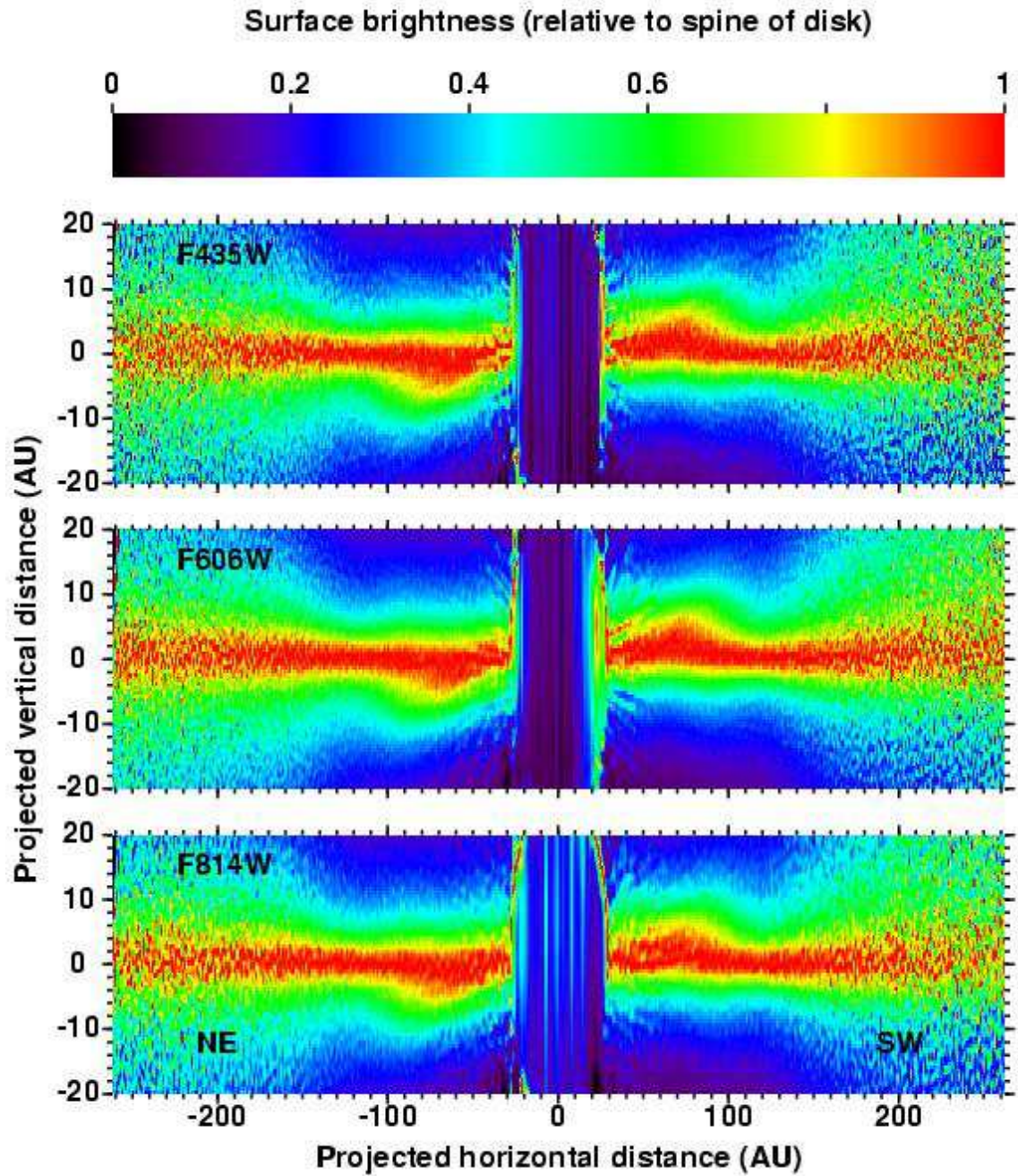}
   \caption{
	Same images shown in Figure~\ref{bvispine} after Lucy--Richardson deconvolution of the ``off-spot'' PSF.  The 
	color-coded isophotes are more similar among these deconvolved filter images than among the convolved images 
	shown in Figure~\ref{bvispine}.  Side effects of the deconvolution process are described in Figure~\ref{bvidisklucy}.
	}
  \label{bvispinelucy}
\end{figure*}

\begin{figure*}[t]
   \epsscale{1.00}\plotone{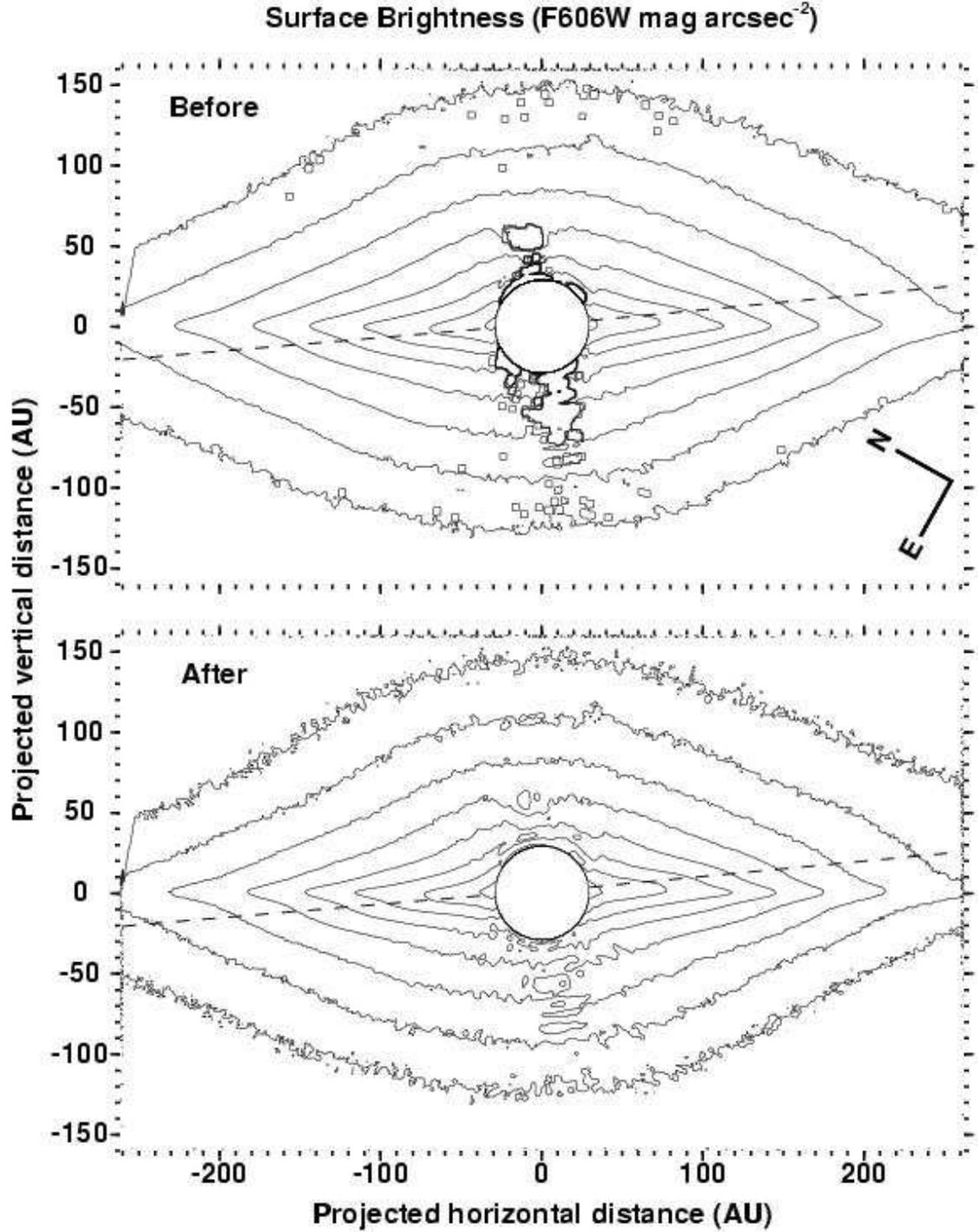}
   \caption{
	Isophotal map of our F606W images of the disk before PSF deconvolution {\it (top panel)} and after PSF deconvolution
	{\it (bottom panel)}.  The images were smoothed with a $9 \times 9$-pixel boxcar.  The interval between isophotes 
	is 1~mag~arcsec$^{-2}$; the outermost isophote in each panel marks a surface brightness of 20~mag~arcsec$^{-2}$.  
	A circular mask of radius 1\farcs5 ($\sim 30$~AU) has been imposed on the innermost region of the disk to reduce 
	confusion from PSF-subtraction residuals.  Each dashed line represents the least-squares fit to the midplane of the 
	corresponding extension of the inner, secondary disk (see \S\ref{mainsec}).
        }
  \label{diskcon}
\end{figure*}

\begin{figure*}[t]
   \epsscale{1.00}\plotone{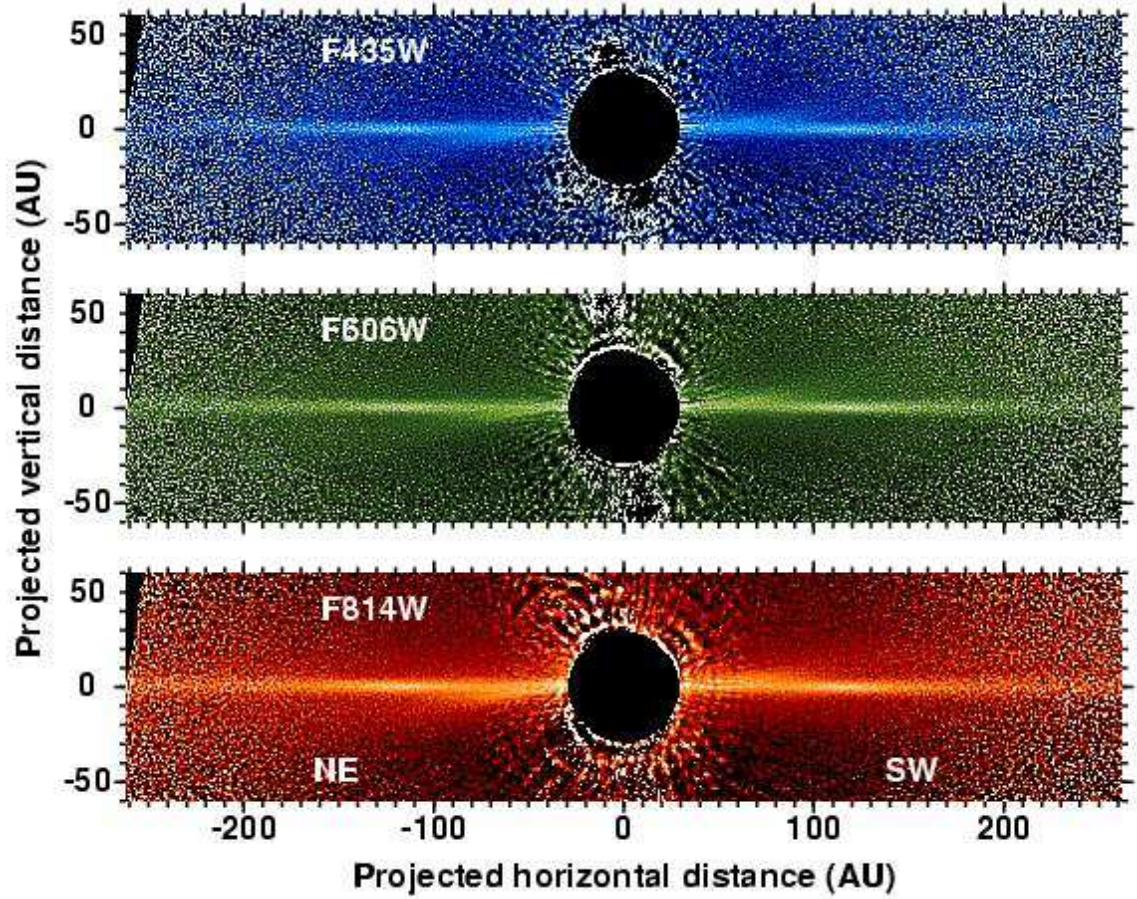}
   \caption{
	Ratios of HRC filter images after and before deconvolution of the off-spot coronagraphic PSF (i.e., images in 
	Figure~\ref{bvidisklucy} divided by corresponding images in Figure~\ref{bvidisk}).  These images accentuate
	the sharply-peaked midplane of the disk and support the notion that the inner warp is a secondary disk, 
	distinct from the main outer disk and inclined from it by $\sim 5^{\circ}$.  A circular mask of radius 
	1\farcs5 ($\sim 30$~AU) has been imposed on the innermost region of the disk to reduce confusion from 
	PSF-subtraction residuals.
        }
  \label{imratio}
\end{figure*}

\begin{figure*}[t]
   \epsscale{0.95}\plotone{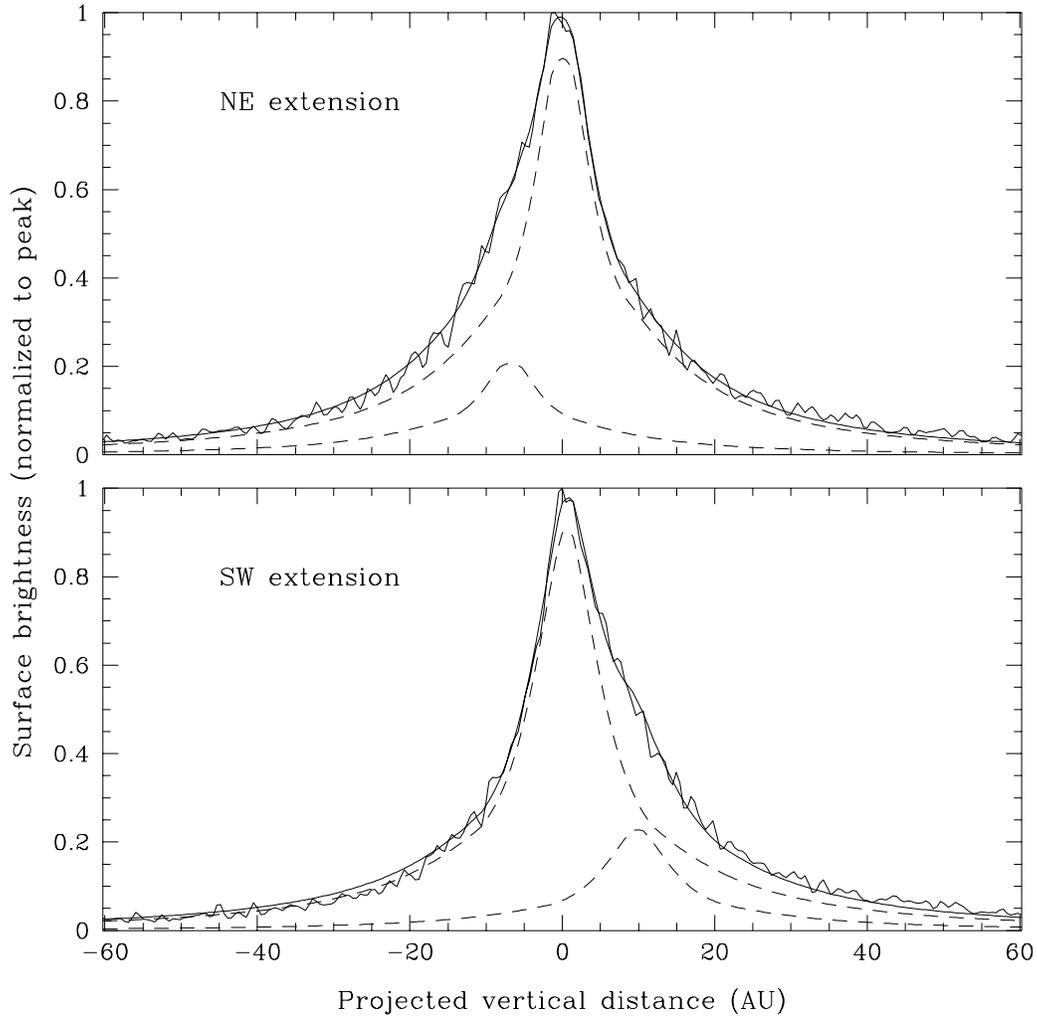}
   \caption{
	The vertical scattered-light profiles (i.e., the scattered-light distributions perpendicular to the midplane) of 
	each extension of the disk at a projected horizontal distance of 100~AU from $\beta$~Pic.  The jagged solid curves
        are the observed profiles extracted from our F606W images after PSF-deconvolution.  Each smooth solid curve is the
        sum of two hybrid-Lorentzian functions (shown as dashed curves; see \S\ref{mainsec}) that best fits the respective observed 
	profile.  The dashed curves represent the contributions of the main and secondary disks to the composite vertical profile.
	The asymmetry of each composite profile and the reversal of this asymmetry in opposing extensions of the disk reflect the 
	``butterfly asymmetry'' characterized by \citet{kal95}.
	}
  \label{vertprof}
\end{figure*}

\begin{figure*}[t]
   \epsscale{0.95}\plotone{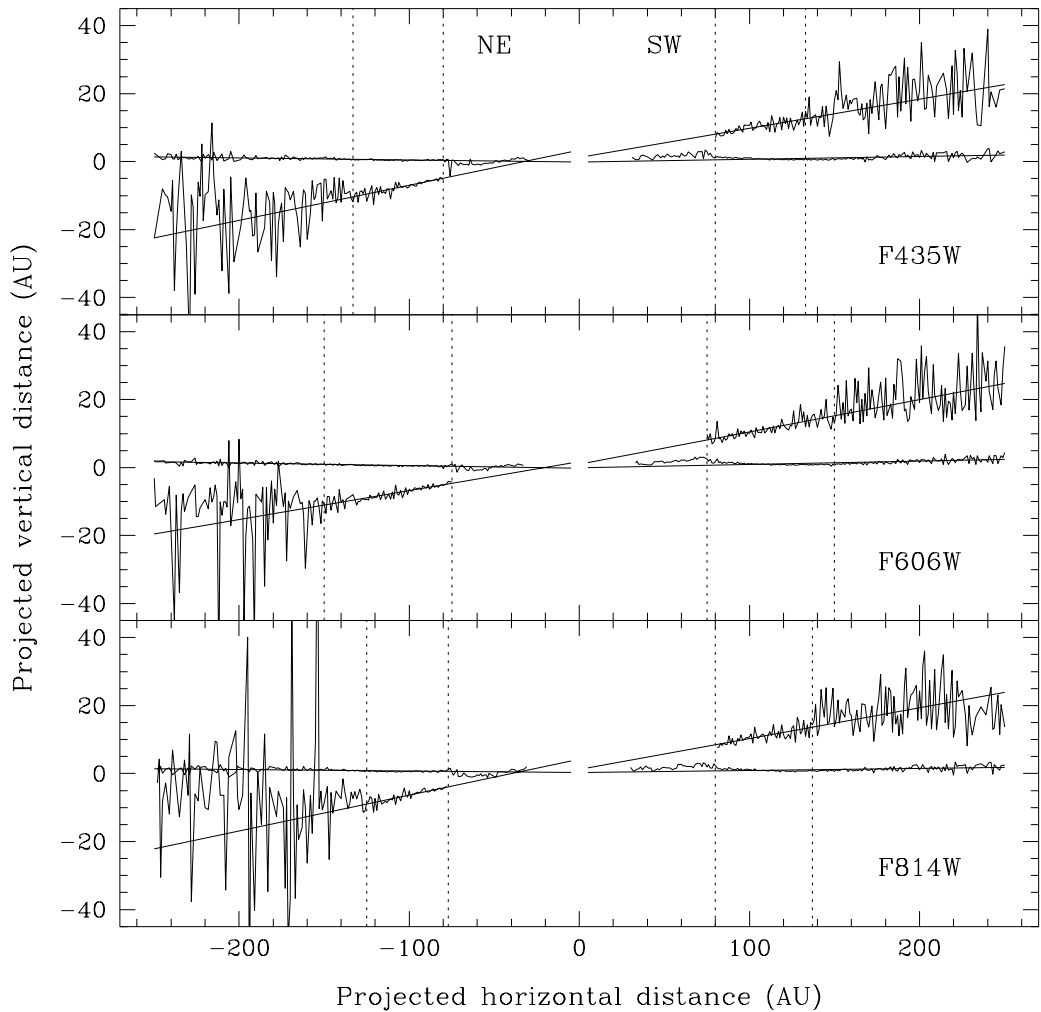}
   \caption{
	Traces of the two ``hybrid-Lorentzian'' components of the composite vertical profiles (e.g., Figure~\ref{vertprof}) at horizontal
	distances of 30--250~AU from $\beta$~Pic.  The vertical dotted lines bound the regions where the composite profiles can be 
	accurately decomposed into two profiles associated with the main (horizontal) and secondary (tilted) disks.  The solid lines
	are least-squares fits to the traces along each disk extension in the regions between 80 and 250~AU for the main disk and 
	between the vertical dotted lines for the secondary disk.  Profile decomposition breaks down within $\sim 80$~AU of the star,
	as evidenced by the large deviations of the main disk components from their respective linear fits.  The noisy traces of the 
	secondary disk within $\sim 80$~AU have been omitted for clarity's sake.
	}
  \label{traceprof}
\end{figure*}

\begin{figure*}[t]
   \epsscale{0.95}\plotone{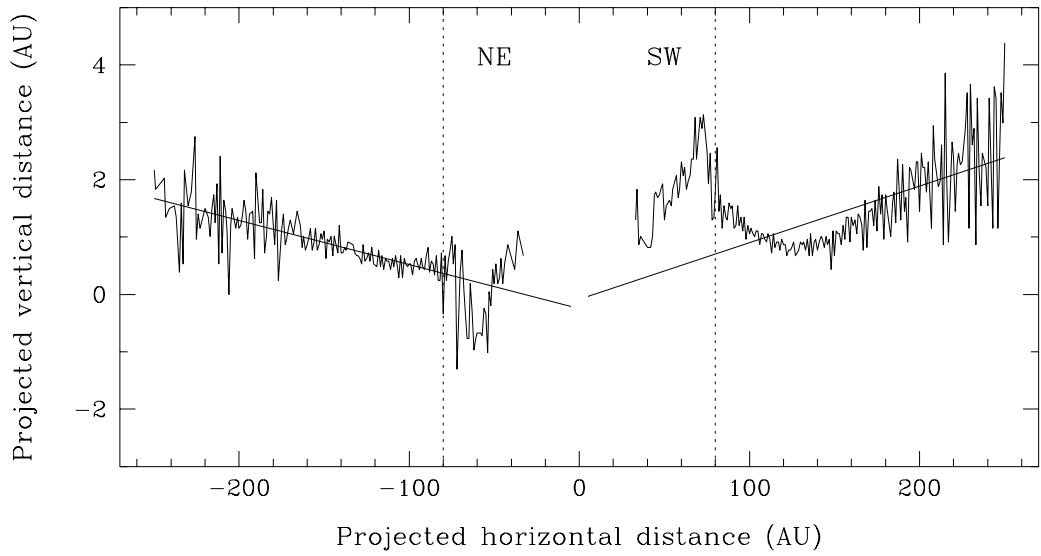}
   \caption{
	Magnified traces of the main disk component observed in our F606W image after PSF deconvolution.  The 
	vertical dotted lines mark the 80~AU boundary within which the composite vertical scattered-light profile
	cannot be credibly decomposed into its main and secondary disk components.  The solid lines are the 
	least-squares fits to the traces along each extension of the main disk at distances of 80--250~AU from 
	$\beta$~Pic.  The opposite slopes of the linear fits reflect a persistent but diminished ``wing-tilt asymmetry,'' 
	first noted by \citet{kal95} at large distances from the star.  The general conformity of the northeast
	extension to the linear fit is not observed in the southwest extension.  The opposing traces obtained from 
	the F435W and F814W images show similar asymmetry; their average vertical positions deviate from the F606W 
	traces by $< 0.5$~AU (i.e., less than one HRC pixel) beyond 80~AU from the star.
        }
  \label{maintrace}
\end{figure*}

\begin{figure*}[t]
   \epsscale{0.95}\plotone{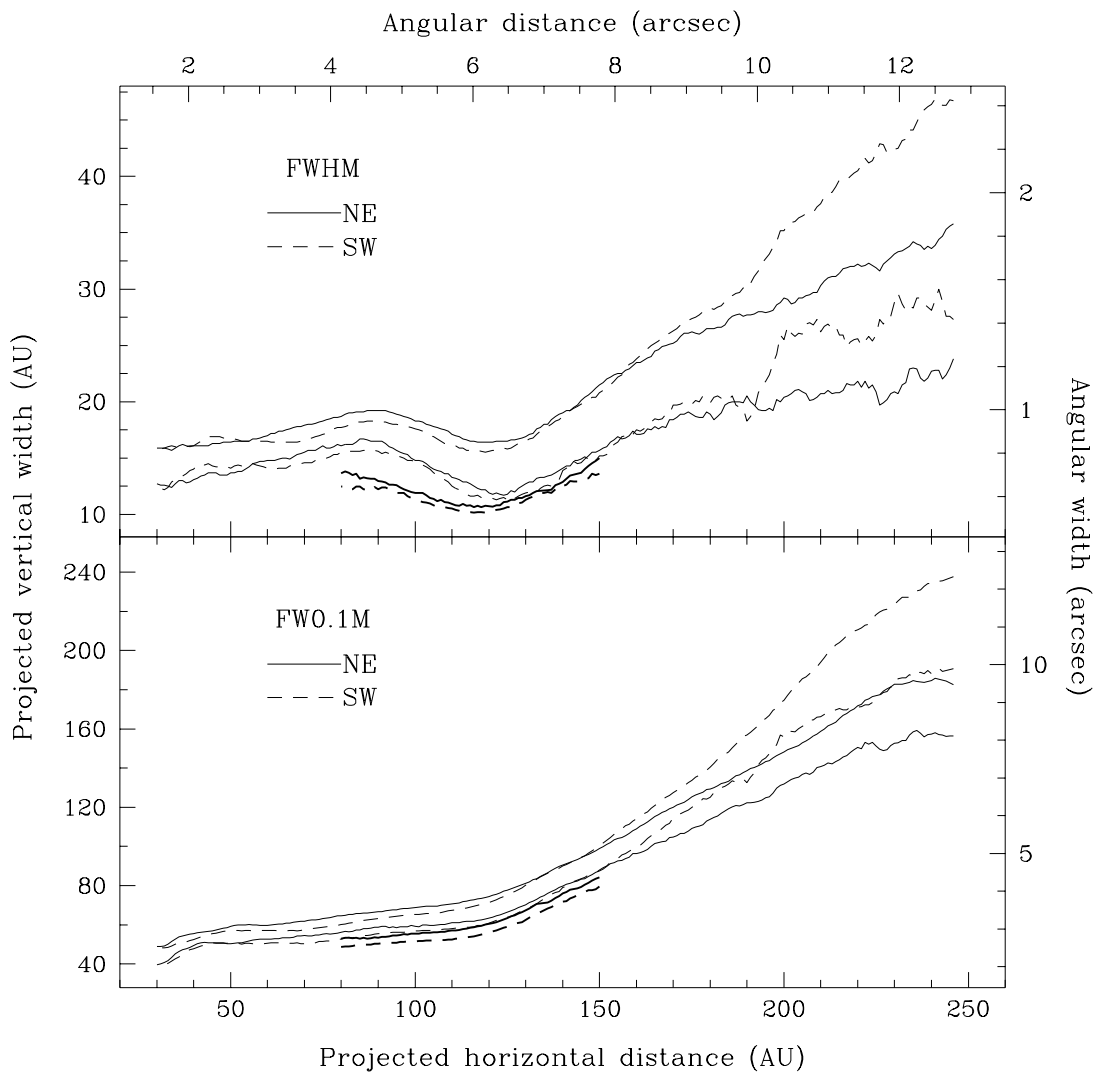}
   \caption{
	The projected vertical width of the composite disk as a function of horizontal distance from $\beta$~Pic.  The 
	curves trace the full widths of each extension of the disk at half {\it (top panel)} and one tenth {\it (bottom 
	panel)} of the maximum (midplane) brightness.  The upper pair of thin solid and dashed curves in each panel show 
	the widths of the northeast and southwest extensions, respectively, measured from the F606W image before 
	deconvolution of the off-spot PSF.  The lower pair of thin curves show the reduced widths after PSF deconvolution.
	The short, thick pair of curves in each panel show the widths of the main component of the disk between 80 and
        150~AU after decomposition and extraction of the tilted secondary disk.  All curves have been smoothed with a 
	9-pixel boxcar.
	}
  \label{diskwidth606}
\end{figure*}

\begin{figure*}[t]
   \epsscale{0.95}\plotone{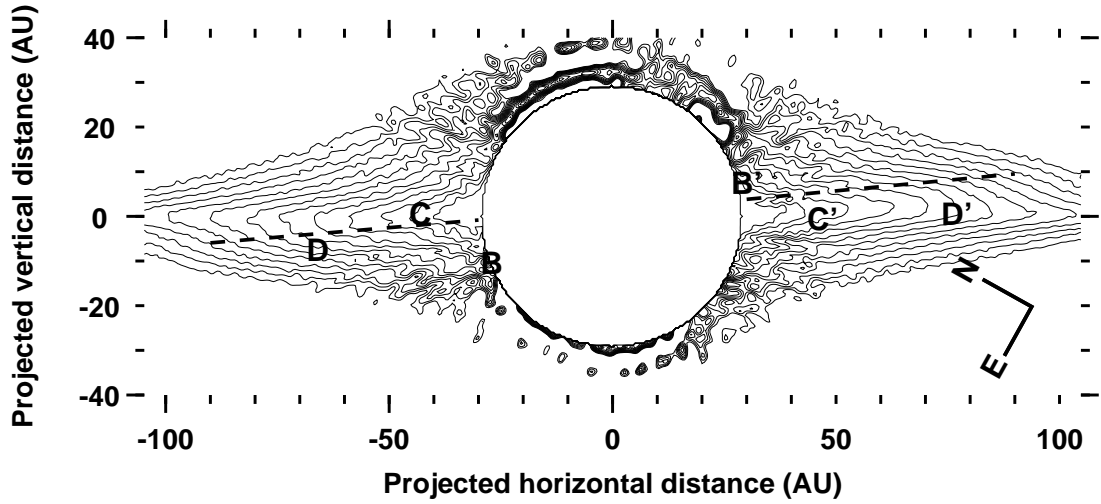}
   \caption{
	Isophotal map of the inner region of the disk obtained from our PSF-deconvolved F606W image.  The data
	have been smoothed with a $3 \times 3$-pixel boxcar.  The isophotes represent surface brightnesses 
	of 10--15.4~mag~arcsec$^{-2}$ at intervals of 0.2~mag~arcsec$^{-2}$.  The dashed lines are the best 
        linear fits to the spine of the secondary disk.  The locations of the diametrically-opposed clumps of 
	$18~\mu$m emission are marked with the letters assigned to them by \citet{wah03}.  The bright clump of
	emission reported by \citet{tel05} is nearly coincident with clump C$'$.  No such clumping is evident 
	in scattered optical light.
	}
  \label{ringcon}
\end{figure*}

\begin{figure*}[t]
   \epsscale{0.95}\plotone{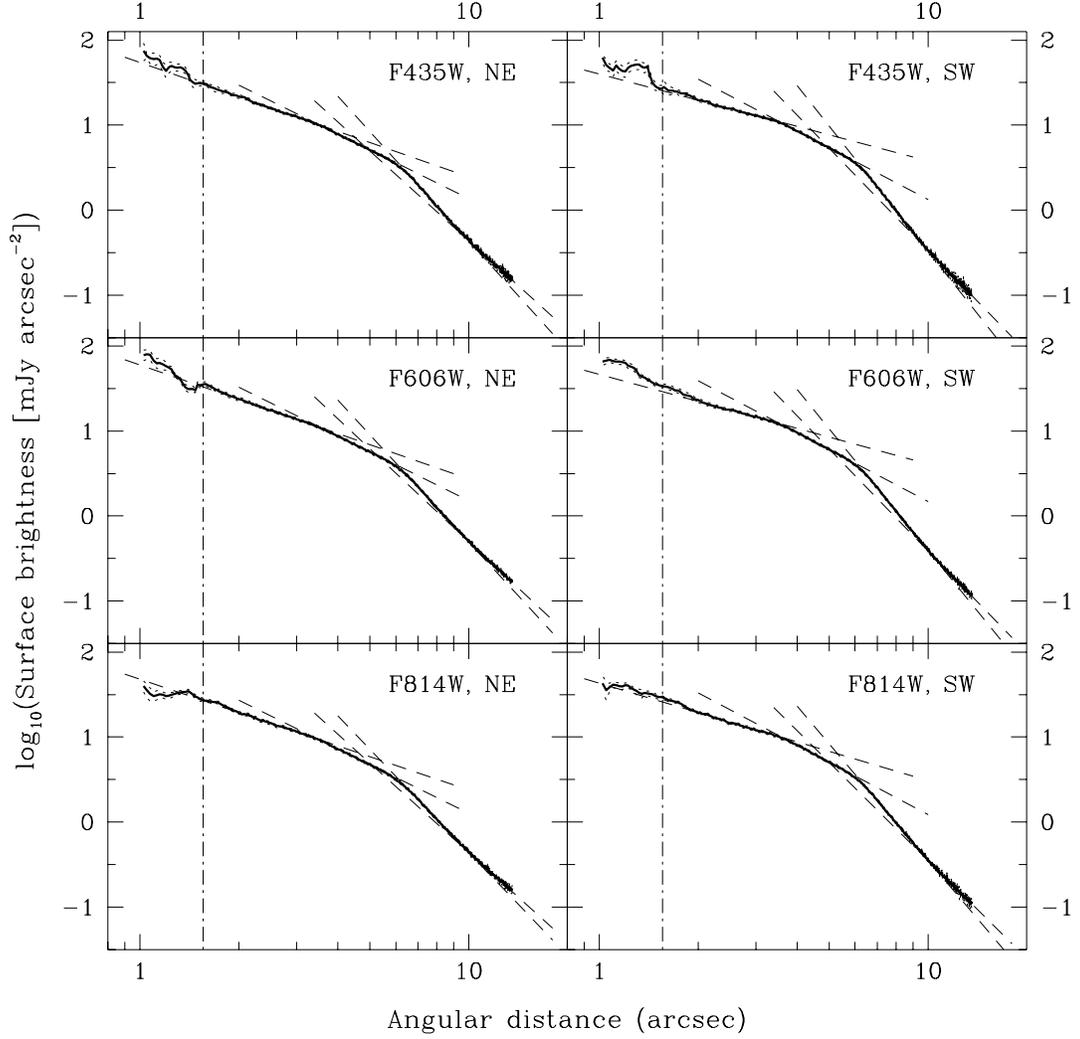}
   \caption{
	Multiband surface brightness profiles measured along the spine of each disk extension before PSF deconvolution.  The
	dotted curves are $\pm 1\sigma$ error profiles derived from the total-error maps of the images (\S\ref{psfsub}) and $\sim 3$\% 
	uncertainty in the photometric calibration.  The dashed lines are least-squares fits to the logarithmic data at angular
	distances of $2''$--3\farcs5, 3\farcs7--5\farcs6, 6\farcs--$10''$, and $10''$--13\farcs4 from $\beta$~Pic.  The slopes
	of these lines (i.e., the indices of the equivalent power-law fits) are given in Table~\ref{powlaw1}.  The vertical 
	lines at $\sim 1$\farcs5 mark the inner limit of credible photometry.
	}
  \label{midpfit}
\end{figure*}

\begin{figure*}[t]
   \epsscale{0.95}\plotone{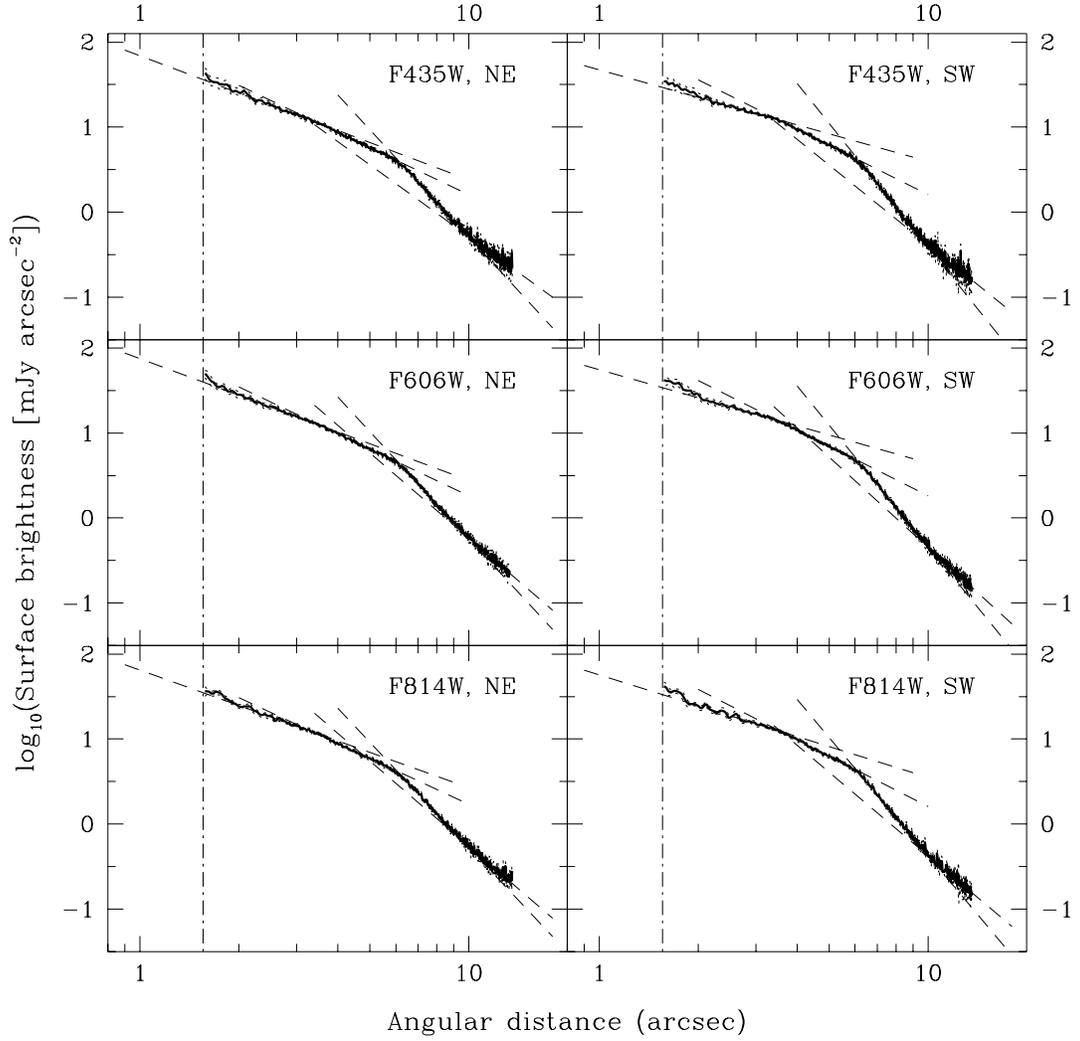}
   \caption{
        Multiband surface brightness profiles measured along the spine of each disk extension after PSF deconvolution.  The
        dotted curves are $\pm 1\sigma$ error profiles obtained by combining in quadrature the $\pm 1\sigma$ errors of the 
	PSF-convolved images (Figure~\ref{midpfit}) and local estimates of the correlated noise imparted by the Lucy--Richardson
	algorithm.  The dashed lines are least-squares fits to the logarithmic data for the regions described in Figure~\ref{midpfit};
	their slopes are given in Table~\ref{powlaw1}.  The vertical lines at $\sim 1$\farcs5 mark the boundary of the circular 
	pixel mask used during PSF deconvolution.
        }
  \label{midpfitlucy}
\end{figure*}

\begin{figure*}[t]
   \epsscale{0.95}\plotone{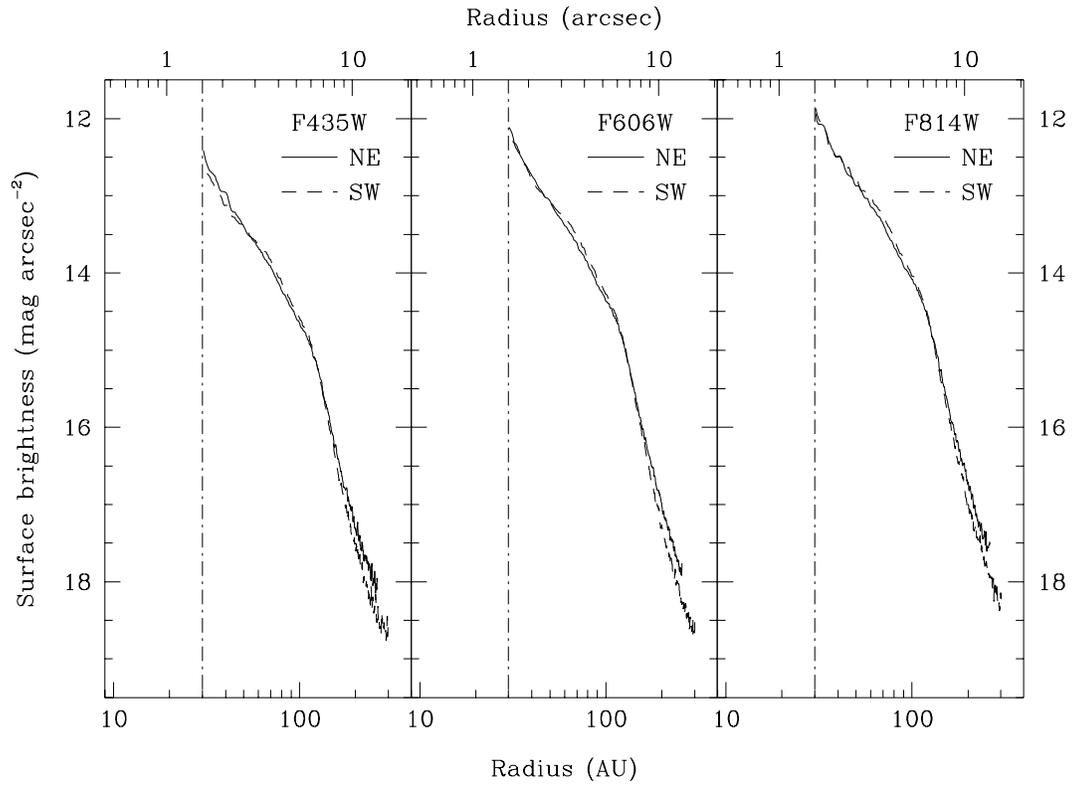}
   \caption{
	Multiband surface brightness profiles measured along the spine of each disk extension after PSF deconvolution.  The
	data have been smoothed with a 5-pixel boxcar to improve clarity.  The dashed lines at 30~AU mark the boundary of the 
	circular pixel mask used during PSF deconvolution.  The uncertainty in the photometric calibration of each bandpass is 
	$\sim 0.03$~mag.
	}
  \label{midpsblucy}
\end{figure*}

\begin{figure*}[t]
   \epsscale{1.00}\plotone{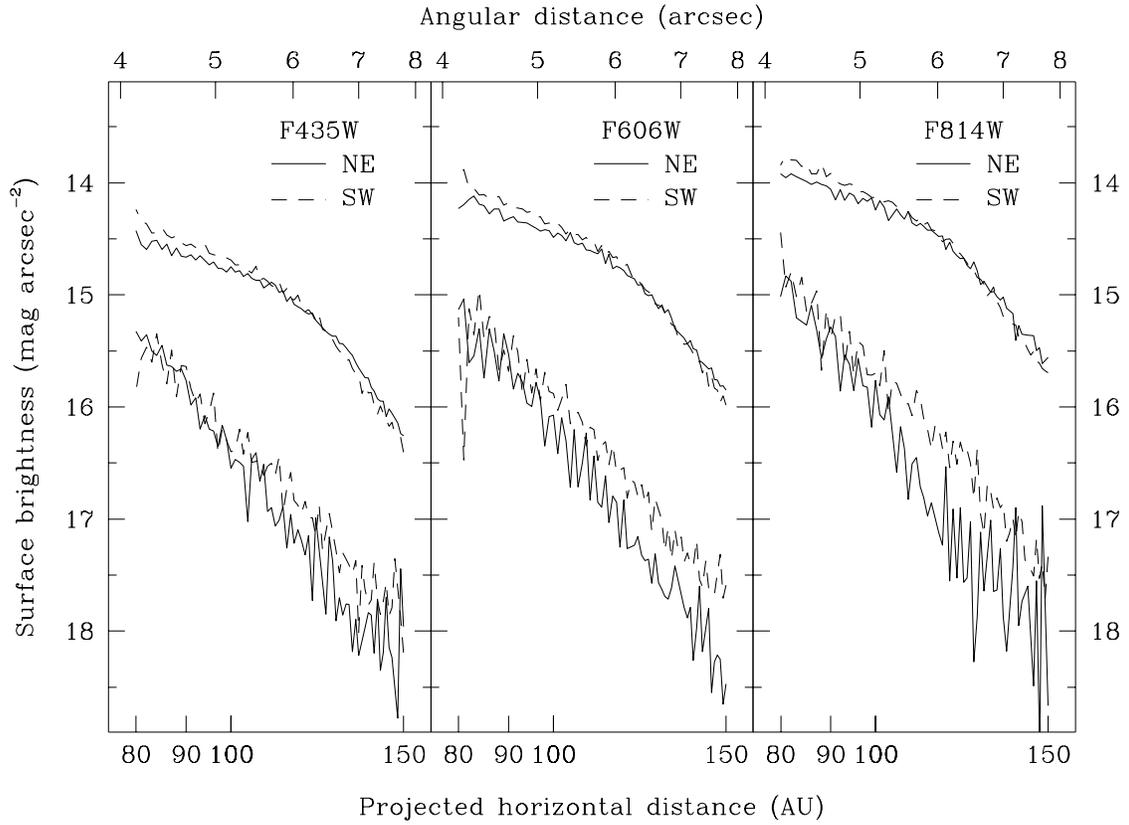}
   \caption{
        Multiband surface brightness profiles of the main {\it (top curves)} and secondary {\it (bottom curves)} disk 
	components, obtained from the PSF-deconvolved images.  The unsmoothed curves trace the maxima of the 
	hybrid-Lorentzian profiles that best fit the vertical profiles of the two components between 80 and 150~AU 
	from $\beta$~Pic (see \S\ref{mainsec}).  The uncertainty in the photometric calibration of each bandpass is 
	$\sim 0.03$~mag.
        }
  \label{mainsecprof}
\end{figure*}

\begin{figure*}[t]
   \epsscale{1.00}\plotone{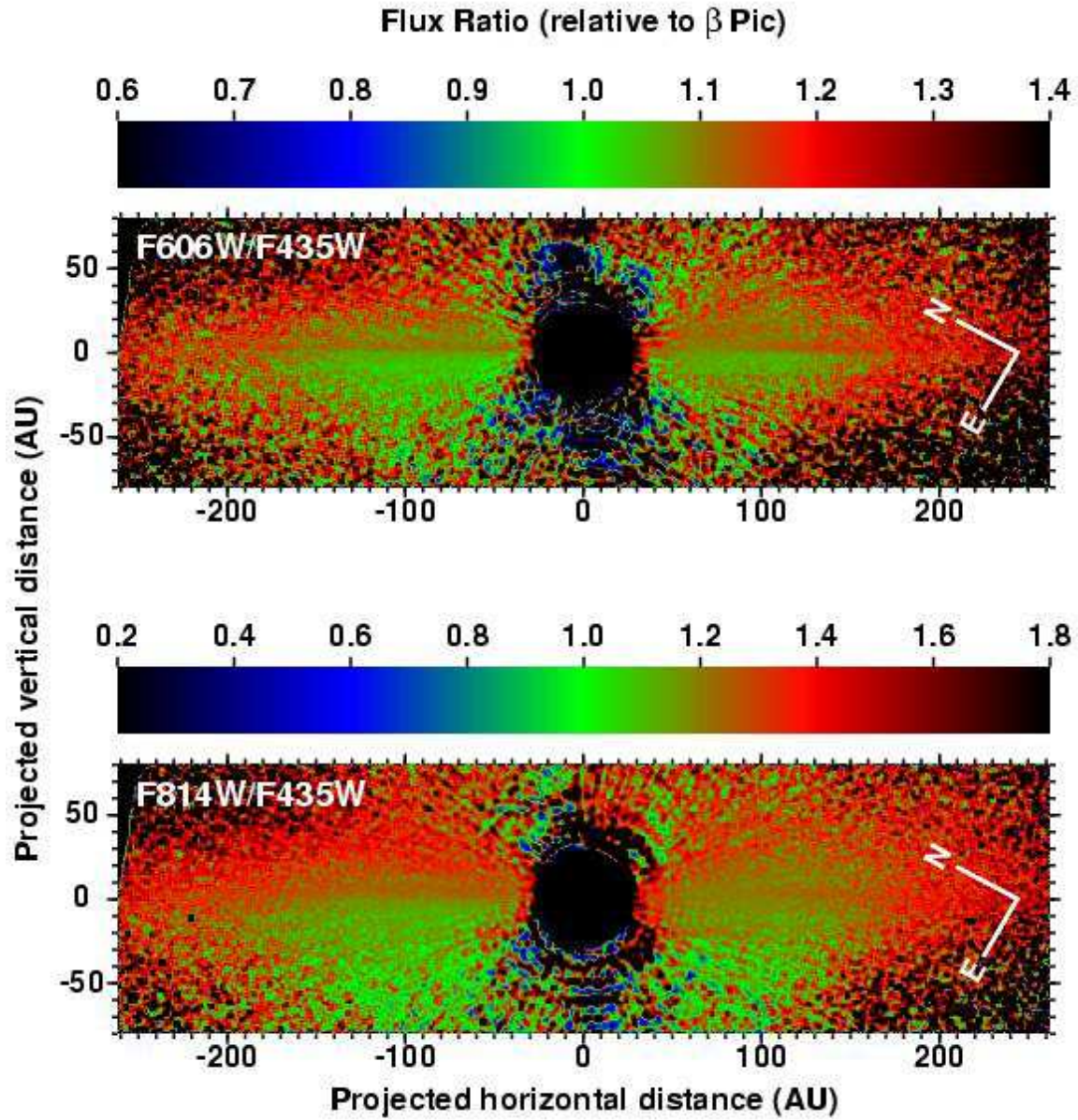}
   \caption{
	F606W/F435W and F814W/F435W ratio images of the composite disk after PSF deconvolution.  The images have been smoothed with a 
	$7 \times 7$ pixel boxcar.  The flux ratios are measured relative to those of the star in the respective bandpasses.  
	Within 150~AU of the star, the uncertainties in the flux ratios range from $\sim 5$--10\% along the spine of the disk to 
	$\sim 25$\% at projected distances of $\pm 50$~AU from the spine.  The F814W/F606W image is not shown, but is qualitatively
	similar to the F606W/F435W and F814W/F435W images.  
	}
  \label{colors2d}
\end{figure*}

Figure~\ref{colors2d} shows the F606W/F435W and F814W/F435W flux ratios of the disk obtained from the PSF-deconvolved images shown in 
Figure~\ref{bvidisklucy}.  Because the images in Figure~\ref{bvidisklucy} are separately normalized by the stellar flux in each bandpass, 
the images in Figure~\ref{colors2d} represent the flux ratios of the disk relative to those of the star.  Consequently, regions of the 
disk having a flux ratio of 1.0 exhibit the same $m_{\rm F435W}$--$m_{\rm F606W}$ and $m_{\rm F435W}$--$m_{\rm F814W}$ colors (hereafter 
denoted F435W--F606W and F435W--F814W) as $\beta$~Pic itself.  Although the flux ratios along the projected minor axis of 
the disk are obscured by the residual artifacts of the PSF subtraction, those along the projected major axis are clear and uncorrupted.
The images show that the flux ratios are asymmetric about both projected axes, with the largest ratios appearing in the north-northeast 
quadrant of the disk.  Within $\sim 150$~AU of the star, the F606W/F435W and F814W/F435W ratios on the southeast side of the disk are 
$\sim 1.0$--1.1 times those of the star, while the ratios on the northwest side are $\sim 1.2$--1.4 times those of the star.  The 
uncertainties of these ratios range from $\sim 5$--10\% along the spine to $\sim 25$\% at projected vertical distances of $\pm 50$~AU from 
the spine.  Beyond $\sim 150$~AU from the star, the color asymmetry between the two sides diminishes substantially, but the uncertainties 
away from the spine are comparatively large ($\sim 25$--75\%).

Figures~\ref{diskcon} and \ref{colors2d} show that the northwest side of the disk within $\sim 150$~AU of $\beta$~Pic is both brighter and 
redder than its southeast counterpart.  If the brightness asymmetry indicates that the side of the disk nearer to Earth is tipped 
northwesterly (\S3.1), then the color asymmetry within $\sim 150$~AU indicates that the ensemble of dust grains in that region becomes 
increasingly forward-scattering with increasing wavelength.  This behavior is inconsistent with the expected scattering properties of both 
compact and porous interstellar grains \citep{dra84,wol98,vos05}, so its cause is not evident.  

The vertical asymmetry of the composite disk's colors is strikingly well aligned with the spine of the disk, which appears prominently 
in both panels of Figure~\ref{colors2d}.  As the S/N ratios of our flux measurements are largest along the spine, we hereafter restrict 
our analysis of the disk's colors to that region in order to assess their dependence on the horizontal projected distance from the star.  
Focusing on the spine also allows us to assess the impact of PSF convolution on the colors measured along this narrow feature of the 
composite disk.

\subsubsection{Colors along the Spine of the Composite Disk\label{compositecolor}}

\begin{figure*}[t]
   \epsscale{1.00}\plotone{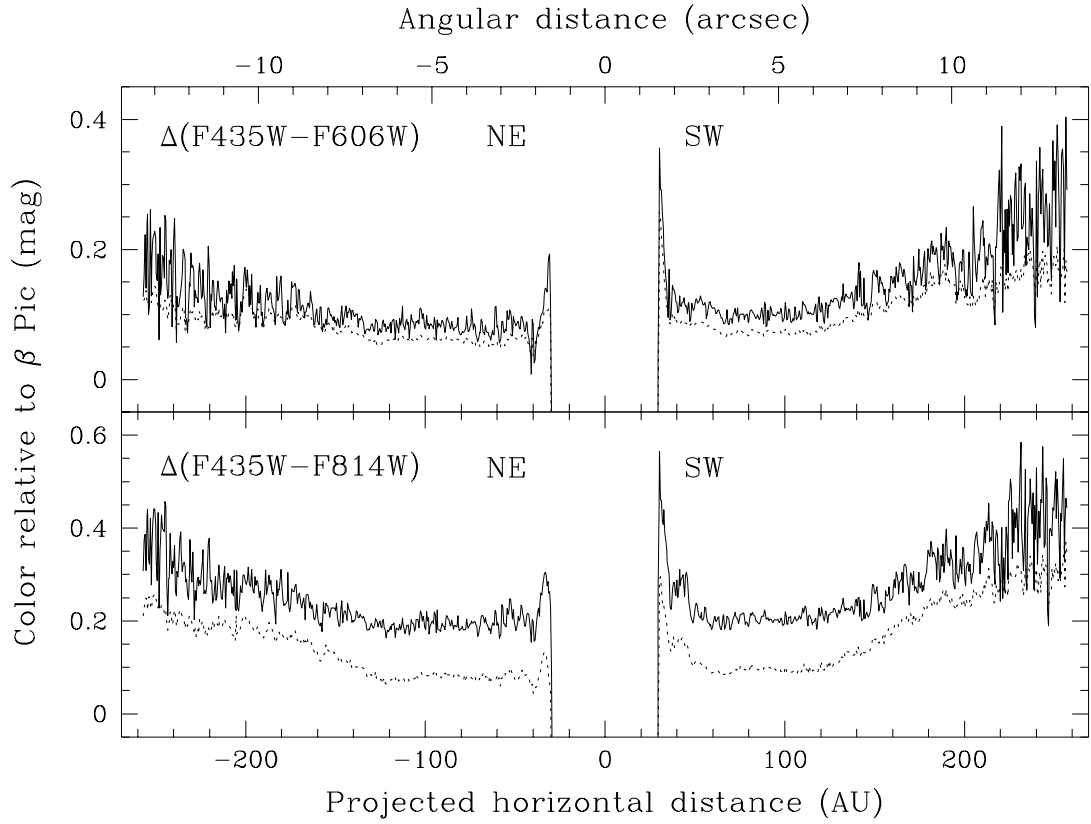}
   \caption{
	F435W--F606W and F435W--F814W colors of the composite disk relative to those of $\beta$~Pic, obtained before 
	{\it (dotted curves)} and after {\it (solid curves)} PSF deconvolution.  The disk's colors were measured along the spines 
	of each extension.  The ratios of the filter images were smoothed with a $7 \times 7$-pixel boxcar for improved clarity.  
	The uncertainties of both colors before PSF deconvolution are $\sim 3$\% at 40--150~AU and $\sim 8$\% at 150--250~AU, 
	whereas those after PSF deconvolution are $\sim 8$\% at 40--150~AU and $\sim 23$\% at 150--250~AU.
        }
  \label{colors}
\end{figure*}

Figure~\ref{colors} shows the F435W--F606W and F435W--F814W colors of the composite disk relative to those of $\beta$~Pic, before and after 
deconvolution of the off-spot PSF.  We measured the colors along the spines of each disk extension after smoothing the F606W/F435W and 
F814W/F435W ratio images with a $7 \times 7$-pixel boxcar.  Although the PSF-convolved colors are less accurate than their deconvolved
counterparts, their uncertainties are more suitably compared with those from previous studies in which PSF deconvolution was not performed.
The average uncertainties of both colors before PSF deconvolution are $\sim 3$\% 
at 40--150~AU and $\sim 8$\% at 150--250~AU, i.e., 3--10 times better than those obtained in previous studies \citep{par87,lec93}.  PSF 
deconvolution increases the differences between the F435W--F606W and F435W--F814W colors of the disk and star by $\sim 0.03$ and $\sim 0.1$~mag,
respectively, at the cost of increased uncertainty.  The relatively large increase in $\Delta$(F435W--F814W) is due to the correction of 
blurring from both the instrumental PSF and the HRC's red-halo anomaly (\S\ref{psfdecon}).  The general similarity of the respective solid and 
dashed curves in Figure~\ref{colors} shows that the color variations as a function of distance from $\beta$~Pic are not significantly affected 
by PSF deconvolution.  This condition is consistent with the practically invariant power-law indices that parametrize the midplane 
surface-brightness profiles in regions 1--3 of the composite disk before and after PSF deconvolution (Table~\ref{powlaw1}).  

The PSF-deconvolved curves in Figure~\ref{colors} show that the F435W--F606W and F435W--F814W colors of the disk are $\sim 0.1$~mag and 
$\sim 0.2$~mag redder, respectively, than those of $\beta$~Pic at distances of $\sim 40$--120~AU from the star.  These color excesses are nearly 
constant within this region and have average uncertainties of $\sim 8$\%.  The F435W--F606W and F435W--F814W excesses respectively increase to 
$\sim 0.2$~mag and $\sim 0.35$~mag at 250~AU in the northeast extension and to $\sim 0.25$~mag and $\sim 0.4$~mag at 250~AU in the southwest 
extension.  The average uncertainties of both excesses beyond 150~AU from the star are $\sim 23$\%, i.e., $\sim 0.05$--0.09~mag at 250~AU.
The constancy of F435W--F606W and F435W--F814W within $\sim 120$~AU is consistent with the chromatically invariant power-law indices associated 
with regions~1 and 2 of the composite disk (Table~\ref{powlaw1}). Likewise, the steadily reddening colors from $\sim 120$--250~AU are consistent 
with the inverse proportionality of $\alpha$ and wavelength in region~3 of the disk.

Our results contradict the longstanding notion that $\beta$~Pic's disk scatters visible light neutrally and uniformly.  Its red optical colors
(relative to the star) are generally consistent with the colors of the disks surrounding the B9~{\small V}e star HD~141569A \citep{cla03} 
and the G2~{\small V} star HD~107146 \citep{ard04,ard05}, although those disks do not exhibit spatial color gradients like those seen in 
$\beta$~Pic's disk beyond $\sim 120$~AU.  The relatively red colors of $\beta$~Pic's disk contrast markedly with the relatively blue colors 
of the edge-on disk surrounding AU~Microscopii \citep{kri05a}, an M dwarf in a co-moving group of stars that includes $\beta$~Pic \citep{bar99,
zuc01mg}.  In \S\ref{aumic}, we speculate that the disparate colors of the disks surrounding $\beta$~Pic and AU~Mic may be caused by small 
differences between the minimum sizes of dust grains in the two disks. 

\subsubsection{Colors along the Spines of the Component Disks\label{componentcolor}}

\begin{figure*}[t]
   \epsscale{1.00}\plotone{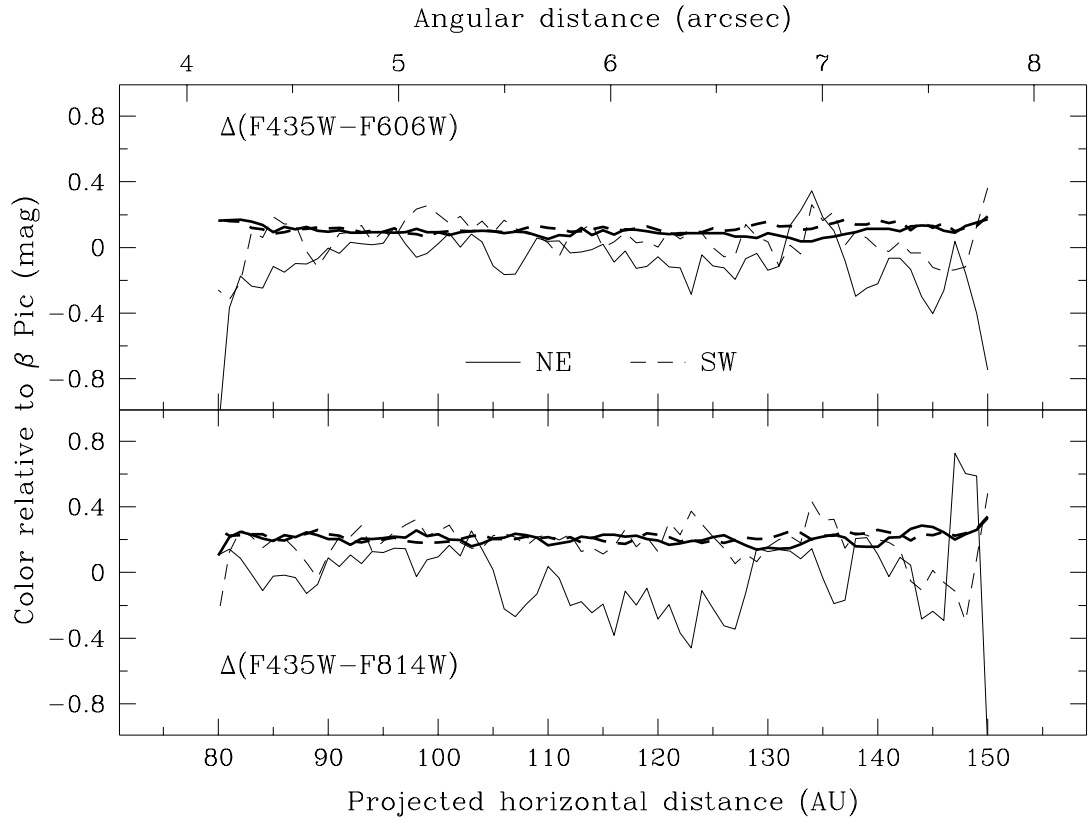}
   \caption{
        F435W--F606W and F435W--F814W colors of the northeast (solid curves) and southwest (dashed curves) extensions of the 
	main (thick curves) and secondary (thin curves) disks, relative to the colors of $\beta$~Pic.  The colors are determined from
	the PSF-deconvolved images using the maxima of the hybrid-Lorentzian profiles fitted to the main and secondary disks 80--150~AU 
	from the star (\S\ref{mainsec}).  The data have been smoothed with a 3-pixel boxcar for clarity.  The RMS deviations of the 
	colors in the main and secondary disks are 0.02~mag and 0.1--0.2~mag, respectively.
        }
  \label{compcols}
\end{figure*}

Figure~\ref{compcols} shows the F435W--F606W and F435W--F814W colors of each extension of the main and secondary disks in the region 80--150~AU from 
$\beta$~Pic.  The colors are derived from the maxima of the hybrid-Lorentzian profiles that best fitted these components of the vertical-scattered
light profiles of the composite disk (\S\ref{mainsec}).  Because the colors depend upon the robustness of our fitting algorithm at each position 
along the spine of the composite disk, their uncertainties cannot be analytically determined.  We estimate these uncertainties from the 
root-mean-squared (RMS) deviations computed for each disk extension.  For the main disk, $\sigma_{\rm (F435W-F606W)} \approx 
\sigma_{\rm (F435W-F814W)} \approx 0.02$ over both extensions.  For the secondary disk, $\sigma_{\rm (F435W-F606W)} \approx 0.1$ over both 
extensions, and $\sigma_{\rm (F435W-F814W)} \approx 0.1$ and 0.2 over the southwest and northeast extensions, respectively.

The colors of the main disk conform to those of the composite disk between 80 and 150~AU, as expected from its nearly fivefold superiority in
brightness over the secondary disk.  The colors of the secondary disk's southwest extension also match those of the main disk, but its northeast 
extension appears bluer than the main disk by $\sim 0.1$--0.2~mag in F435W--F606W and $\sim 0.1$--0.4~mag in F435W--F814W.  Moreover, the 
F435W--F814W color of this extension apparently becomes bluer with increasing distance from the star, as previously suggested by its steepening 
midplane surface-brightness gradient (\S\ref{componentprof} and Table~\ref{powlaw2}).  However, the relatively large uncertainties in this region 
of the disk preclude a definitive assessment of the colors and trends of the secondary disk.  Within our estimated errors, the colors of the 
secondary disk are also consistent with being neutral and constant between 80 and 150~AU.

\section{Discussion\label{discussion}}

\subsection{Impact of PSF deconvolution on observations and models\label{impact}}

Our analysis demonstrates that PSF deconvolution is an important step toward accurately characterizing thin debris disks viewed edge-on or, more 
generally, less-inclined disks having narrow or compact structural elements.  It is especially important for multiband imaging studies designed 
to map the chromatic dependencies of a disk's morphology and surface brightness.  For example, PSF deconvolution dispels the illusion that 
$\beta$~Pic's disk thickens with increasing wavelength (Figures~\ref{bvispine} and \ref{bvispinelucy}) by restoring the displaced red flux to 
its original location along the disk's spine.  This restoration of the true surface-brightness distribution contributes largely to our observation
that the main disk component is significantly redder than previously determined from ground-based measurements \citep{par87,lec93}.  It also 
provides refined empirical constraints for past and future dynamical models that alternatively view the inner warp as a uniform, propagating 
deformation of a single disk or as two distinctly separate, inclined disks \citep{mou97b,aug01}.

The improved measurements of the disk's morphology and surface brightness obtained from our PSF deconvolved images provide a basis for improved 
scattered-light models of $\beta$~Pic's disk.  However, deriving the spatial distribution and scattering properties of dust in an edge-on disk is 
complicated by the integrated effects of variations in both characteristics along the line of sight.  This complication is compounded by the many 
asymmetries between the opposing extensions of $\beta$~Pic's disk and by the existence of an inclined secondary disk.  Consequently, three-dimensional 
scattering models based on axisymmetric dust distributions -- like those previously developed for the disks around $\beta$~Pic \citep{art89,kal95} and 
AU~Mic \citep{kri05a} -- have limited utility in the face of the morphological and photometric complexity of $\beta$~Pic's disk revealed by our 
HRC images.  

Multicomponent, nonaxisymmetric models are clearly needed to reproduce the complicated scattered-light distribution associated with $\beta$~Pic's 
disk.  Development of such models is beyond the scope of this paper, but we identify several issues raised by our observations that should be 
addressed by future modelling efforts:

\begin{enumerate}
\item
The horizontal surface-brightness profiles of the main disk within $\sim 150$~AU are sensitive to contamination by flux from the secondary disk.
Models of the dust distribution within either disk should be based on deconvolved and deblended components of scattered light from both disks.
\item
The spines of the opposing extensions of the secondary disk are not collinear.  While the misaligned spines of the main disk \citep[dubbed the 
``wing-tilt'' asymmetry by][]{kal95} are attributed to forward scattering from a slightly inclined disk, those of the secondary disk are not as 
easily explained.  The possible optical and/or dynamical causes of this phenomenon need to be explored.
\item 
The clumpiness noted in mid-infrared images of the disk is not observed in optical scattered light, even though the grains believed responsible 
for the clumps of mid-infrared emission \citep{oka04,tel05} have sizes commensurate with efficient scattering of optical light.  This apparent 
incompatibility may indicate that the scattered light and thermal emission come from different regions along the line of sight \citep{pan97}.  
If so, these disparate observations may constrain the mean scattering properties of the grains.
\item 
The optical colors of the composite disk are asymmetric about both projected axes of the disk.  The generally redder and brighter appearance 
of the northwest half of the disk is inconsistent with the expectation that the dust grains scatter red light more isotropically than blue light.
This peculiar behavior may place interesting constraints on the sizes and compositions of the grains throughout the disk.
\item 
The prominent changes in the horizontal surface-brightness profiles and color gradients at $\sim 115$~AU from $\beta$~Pic suggest that these 
phenomena may have a common cause.  The expected sublimation of water ice from micron-sized grains lying within $\sim 100$~AU of the star
\citep{pan97,li98} is one possible cause, but color variations caused by changes in the grain-size distribution must also be investigated.
\end{enumerate}

\noindent
Although we defer these issues to future theoretical investigations, we partly address the last issue here via a one-dimensional 
analysis of the colors observed along the spines of the component disks.

\subsection{Colors and Properties of the Dust\label{dustprops}}

\citet{li98} successfully modeled the disk's continuum emission from near-infrared to millimeter wavelengths and its $10~\mu$m silicate 
emission feature by assuming that the dust is continually replenished by comets orbiting near, or falling onto, $\beta$~Pic.  In this 
scenario, the cometary dust grains are highly porous aggregates of primitive interstellar dust whose composition, molecular 
structure, and size and spatial distributions are altered by the stellar radiation environments into which they are sputtered and 
transported.
\citet{li98} did not apply this dust model to the disk's appearance in scattered light 
because the scattering properties of highly porous grains with appropriate compositions had not been fully determined.  Subsequent 
modeling of these properties \citep{wol98,vos05} now enables us to examine the porous-grain model from the perspective of the disk's
optical colors.  

We begin by deriving a parameter that, in the absence of a detailed three-dimensional model of the disk, links the optical properties of a 
variety of grains with the colors observed along the spine of the disk.  We apply this parameter first to the colors of the composite disk 
(Figure~\ref{colors}), as these colors are largely attributable to the main disk and their uncertainties are well established.  Afterwards, we 
examine the less-certain colors measured for the secondary disk (Figure~\ref{compcols}).

\subsubsection{Effective Scattering Efficiency of Midplane Grains\label{qeff}}

The intensity of singly-scattered light measured along the midplane of an optically thin, edge-on disk is 

\begin{equation}
I(\epsilon) = \int n(r)~\sigma_{\rm sca}~\Phi(\theta)~F_0 \left(\frac{r}{r_0}\right)^{-2}~dx,
\end{equation}

\noindent
where $\epsilon$ is the angular distance from the star, $n(r)$ is the number density of grains at a distance $r$ from the star, $\sigma_{\rm sca}$
is the scattering cross section of grains of a given size and composition, $\Phi(\theta)$ is the phase function of the scattering angle $\theta$, 
$F_0$ is the stellar flux at an arbitrary radius $r_0$, and $x$ traverses the disk along the line of sight.  The complicated and asymmetric surface 
brightness profiles of $\beta$~Pic's disk suggest that $n(r)$ cannot be expressed as a single analytic function throughout the disk.  However, 
constraining $n(r)$ within each region of each disk extension requires a complex modelling effort that is beyond the scope of this paper.   

For a less rigorous analysis of the disk's colors, we approximate the surface brightness and number density profiles along the midplanes of both 
extensions with single radial power laws having fixed indices $-\alpha$ and $-\nu$, respectively.  Because the outer radius of the disk is $\gtrsim 
1800$~AU \citep{lar01}, our lines of sight mostly traverse regions of the disk that lie well beyond the projected distances at which we see prominent
inflections in the surface brightness profiles.  The effective index $\alpha$ should therefore be biased toward the indices observed in the outer 
regions of the disk, i.e., $\alpha \approx 4$ \citep{smi84,kal95}.  \citet{nak90} determined that $\nu = \alpha - 1$ if the observed scattered 
light originates mostly from dust along those parts of line of sight near $\beta$~Pic.  This condition is valid unless the dust grains are strongly 
forward-scattering.  Some models indicate that the scattered light is dominated by such grains \citep{pan97}, but the relationships between the
sizes, compositions, and optical properties of the grains remain uncertain.  Assuming for simplicity that the disk comprises grains that are at most
moderately forward-scattering, we obtain

\begin{equation}
n(r) = n_0 \left(\frac{r}{r_0}\right)^{-3}.
\end{equation} 

Adopting the disk geometry portrayed in Figure~1 of \citet{bui86}, we rewrite Equation~(1) as

\begin{equation}
I(\epsilon) = \frac{F_0~n_0~r_0^5}{R^4~{\rm sin}^4\epsilon}~\int_{\theta_0}^{\pi - \theta_0} \sigma_{\rm sca}~\Phi(\theta)~{\rm sin}^3\theta~d\theta,
\end{equation}

\noindent
where $R$ is the distance to $\beta$~Pic, and $\theta_0$ is the scattering angle at the outer edge of the disk.  The radius of the disk greatly 
exceeds the FOV of our images, so $\theta_0 \approx 0$.  The integral is then solely dependent upon the scattering properties of the grains, so 
we define the grains' effective scattering efficiency as 

\begin{equation}
Q_{\rm eff} = \int_{0}^{\pi} Q_{\rm sca}~\Phi(\theta)~{\rm sin}^3\theta~d\theta,
\end{equation}

\noindent
where $Q_{\rm sca} = \sigma_{\rm sca}/\pi a^2$, and $a$ is the grain size (radius).  We use a common scattering phase function for small 
grains, 

\begin{equation}
\Phi(\theta) = \frac{1-g^2}{4\pi~(1 + g^2 - 2g~{\rm cos}~\theta)^{3/2}},
\end{equation}

\noindent
where $g$ is the scattering asymmetry parameter defined by

\begin{equation}
g = \langle cos~\theta \rangle = \int_{4\pi} \Phi(\theta)~{\rm cos}~\theta~d\Omega
\end{equation}

\noindent
and $d\Omega$ is the unit solid angle \citep{hen41}.  Values of $g = -1$, 0, and 1 correspond respectively to backward, isotropic, and forward
scattering.

Substituting Equation~(5) into Equation~(4) and evaluating the integral, we obtain

\begin{equation}
Q_{\rm eff} = \frac{Q_{\rm sca}}{3\pi}~(1-g^2).
\end{equation}

\noindent
This result indicates that our view of the edge-on disk is produced by mostly isotropic scatterers, which conforms to our initial assumption 
that the grains in the disk are not strongly forward-scattering.  Its analytic form is a fortunate consequence of our choice of $\nu = 3$,
but the bias against forward-scattering grains is maintained for any choice of $\nu > 1$.  Both $Q_{\rm sca}$ and $g$ are functions of the 
dimensionless parameter $x = 2 \pi a / \lambda$, where $\lambda$ is the wavelength of the scattered light.  The mean $Q_{\rm eff}$ for grains 
of size $a$, weighted by the stellar flux in a given bandpass, is

\begin{equation}
\langle Q_{\rm eff} \rangle = \frac{\int Q_{\rm eff} F_\lambda T_\lambda~d\lambda}{\int F_\lambda T_\lambda~d\lambda},
\end{equation}

\noindent
where $F_\lambda$ is the flux spectrum of $\beta$~Pic and $T_\lambda$ is the filter transmission profile.  We now compare 
$\langle Q_{\rm eff} \rangle$ for grains of various compositions and porosities in the ACS $B$, Broad-$V$, and Broad-$I$ bands.

\subsubsection{Dust Grains within the Ice-Sublimation Zone\label{innerdisk}}

\begin{figure*}[t]
   \epsscale{1.00}\plotone{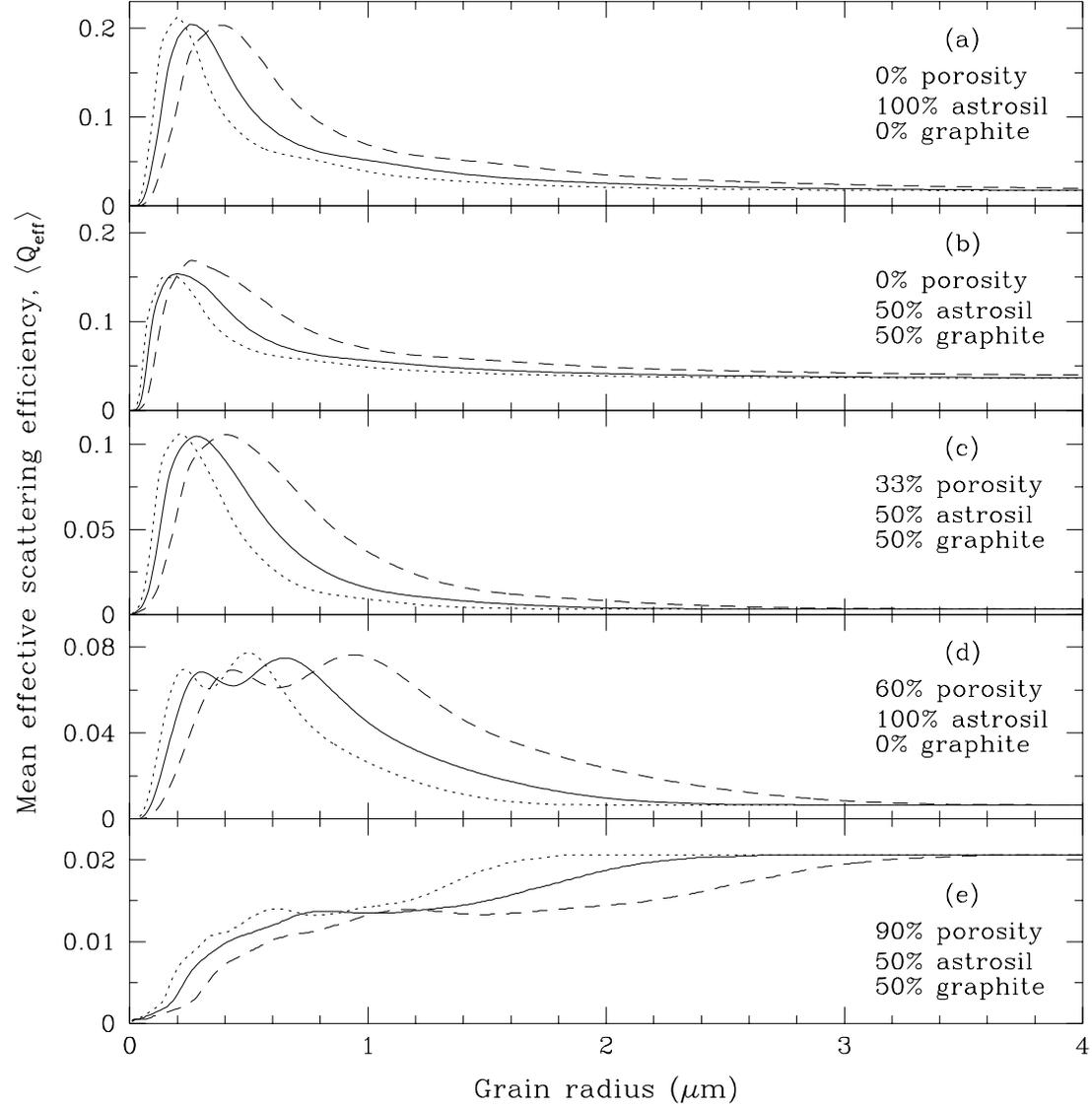}
   \caption{
	$\langle Q_{\rm eff} \rangle$ versus grain radius for F435W {\it (dotted curves)}, F606W {\it (solid curves)}, F814W {\it (dashed curves)} 
	and five combinations of porosity and composition: (a) compact grains of pure astronomical silicate \citep[``astrosil,''][]{dra84};
	(b) compact grains with equal amounts of astrosil and graphite \citep{dra84}; (c) 33\% porous grains with equal numbers of astrosil 
	and graphite inclusions \citep{vos05}; (d) 60\% porous grains of pure astrosil \citep{wol98}; and (e) 90\% porous grains with equal 
	numbers of astrosil and graphite inclusions \citep{vos05}.  The curves are computed for an assumed grain number density of $n(r)
        \propto r^{-3}$.
        }
  \label{qmean}
\end{figure*}

Figure~\ref{qmean} shows $\langle Q_{\rm eff} \rangle$ as a function of $a$ for the F435W, F606W, and F814W filters and five combinations of 
porosity and composition.  None of these combinations includes icy mantles, so they are applicable only to the region of the disk within the ice 
sublimation limit \citep[$\lesssim 100$~AU for micron-sized grains;][]{pan97,li98}.  Figures~\ref{qmean}a and \ref{qmean}b depict $\langle Q_{\rm eff} 
\rangle$ for compact (0\% porosity) grains composed, respectively, of pure astronomical silicate (``astrosil'') and equal parts astrosil and 
amorphous carbon (graphite).  The curves are derived from tabulated values of $Q_{\rm sca}$ and $g$ for each composition based upon the work 
of \citet{dra84}.\footnote{As of this writing, tabulated values of $Q_{\rm sca}$ and $g$ for compact grains of astrosil and graphite are 
available on the World Wide Web at http://www.astro.princeton.edu/$\sim$draine/dust/dust.diel.html.}  Figures~\ref{qmean}c and \ref{qmean}e
show $\langle Q_{\rm eff} \rangle$ for 33\% and 90\% porous grains with equal amounts of astrosil and graphite inclusions, based on the 
values of $Q_{\rm sca}$ and $g$ for such grains computed by \citet{vos05}.  Figure~\ref{qmean}d depicts an intermediate case of 60\% porosity 
and pure astrosil derived from values of $Q_{\rm sca}$ and $g$ computed by \citet{wol98}.   For $x > 25$, we extrapolated the tabulated 
values of $Q_{\rm sca}$ and $g$ for porous grains to the geometric-optics limits of $Q_{\rm sca} \rightarrow 1$ and $g \rightarrow 1$ as 
$x \rightarrow \infty$.  This extrapolation is acceptable because even the most porous grains are expected to be strongly forward-scattering
for $x \gg 25$ (N.~Voshchinnikov, personal communication) and thus, for $\nu = 3$, contribute negligibly to our edge-on view of the disk.

Figures~\ref{qmean}a and b represent the grain characteristics most often invoked when modelling scattered-light images of debris disks.  When 
viewed edge-on, disks comprising compact grains with $a \gg 1~\mu$m exhibit similar scattering efficiencies in F435W, F606W, and F814W, i.e, 
they are neutral, albeit relatively inefficient, scatterers at optical wavelengths.  This condition is the basis of previous assertions that 
the reportedly neutral colors of $\beta$~Pic's disk reflect a minimum grain size of several microns \citep{par87,lec93}.  However, as noted by 
\citet{chi91}, the values of $\langle Q_{\rm eff} \rangle$ for smaller grains ($a \approx 0.2$--$0.3~\mu$m) are also similar among the broad 
optical bands, so some submicron-sized grains are also neutral scatterers at these wavelengths.  Moreover, because small grains likely dominate 
the size distribution of grains in the disk \citep{doh69,li98}, they more strongly influence the overall color of the disk than supermicron-sized
grains.  Nevertheless, the truly red colors of $\beta$~Pic's composite disk (Figure~\ref{colors}) makes the debate over the cause of neutral 
colors in this case irrelevant.

Figures~\ref{qmean}c, d, and e show that increasing porosity has two dramatic effects on $\langle Q_{\rm eff} \rangle$.  First, the pronounced
peak at 0.2--$0.4~\mu$m diminishes rapidly for porosities $\gtrsim 60$\%.  This phenomenon is caused by the similar behavior of $Q_{\rm sca}$ for 
$x \lesssim 10$ as porosity increases \citep{wol98,vos05}.  Second, the peak broadens as it diminishes, causing the ranges of predominantly 
blue-, neutral-, and red-scattering grains to broaden and shift to larger grain sizes.  For 90\% porosity, almost all grains with $a \lesssim 
3~\mu$m scatter light in F435W more efficiently than in F606W and F814W, and grains with $a \gtrsim 3~\mu$m are neutral scatterers.  The values of 
$\langle Q_{\rm eff} \rangle$ for blue-scattering grains with $a \lesssim 1~\mu$m are less than half those of the neutrally-scattering grains, 
but no distribution of sizes for 90\% porous grains will yield a disk with red colors.

To compare the colors of an ensemble of grains having uniform porosity and composition with the observed colors of $\beta$~Pic's disk, we must 
determine for each bandpass the mean scattering cross section of the ensemble using $\langle Q_{\rm eff} \rangle$ and a reasonable grain-size 
distribution.  Applying this quantity to Equation~(3), we express the intensity of scattered-light from the disk relative to the stellar flux 
in a given bandpass as, 

\begin{equation}
\mathcal{I} = \frac{I(\epsilon)}{F_*} =  \frac{r_0^3}{R^2 \rm{sin}^4\epsilon}~\int_{a_{\rm min}}^{a_{\rm max}} \pi a^2 \langle Q_{\rm eff}\rangle~\frac{dn_0}{da}~da,
\end{equation}

\noindent
where $F_* = F_0 r_0^2/R^2$ is the stellar flux measured at Earth, $dn_0/da$ is the grain-size distribution, and $a_{\rm min}$ and $a_{\rm max}$ are 
the minimum and maximum sizes of the grains.  The intrinsic color of the disk between bandpasses 1 and 2 is therefore,

\begin{equation}
m_1 - m_2 = -2.5~{\rm log} \left( \frac{\mathcal{I}_1}{\mathcal{I}_2} \right).
\end{equation}

Figure~\ref{dcolors} shows the simulated F435W--F606W and F435W--F814W colors of the inner disk as functions of minimum grain size for the 
grain compositions and porosities previously considered.  To compute these colors, we assumed the values of $\langle Q_{\rm eff}\rangle$ from 
Figure~\ref{qmean} and a power-law size distribution, $dn_0 \propto a^{-3.5}~da$, commonly attributed to dust produced from planetesimal and 
particle collisions \citep{doh69}.  We computed the integral in Equation~9 for $a_{\rm min} = 0.01$, 0.02, ..., $5.00~\mu$m and an 
arbitrary $a_{\rm max} = 100~\mu$m.\footnote{We repeated this exercise for the porous-grain cases using the grain-mass distribution 
derived by \citet{li98} from their thermal dust model that assumed 97.5\% porosity and $n(r) \propto r^{-1.8}$ within the region 
$46 \leq r \leq 115$~AU \citep[Figure~11c of][]{li98}.  The results were nearly identical to those obtained for the grain-size 
distribution of \citet{doh69}, so we discuss them no further.}  Except for the case of 90\% porosity, the observed colors along the spine
of the composite disk within 120~AU (F435W--F606W~$\approx 0.1$ and F435W--F814W~$\approx 0.2$; Figure~\ref{colors}) are simultaneously reproduced 
by all combinations of porosity and composition when $a_{\rm min} = 0.15$--$0.20~\mu$m.  The 90\% porous grains yield only blue or neutral 
colors for any choice of $a_{\rm min}$ (as presaged by Figure~\ref{qmean}e), and therefore alone fail to explain the observed red colors
of the inner disk.

Our measured colors along the spine of the composite disk within 120~AU are, by themselves, insufficient to constrain tightly the composition 
and porosity of the grains in that region.  However, they do show that very porous grains do not contribute significantly to the integrated 
scattered light along the spine.  This conclusion does not necessarily imply a lack of very porous grains, because the values of $\langle 
Q_{\rm eff}\rangle$ for such grains with $a \lesssim 2~\mu$m are 10--20 times smaller than those of compact and moderately porous grains of similar 
size (Figure~\ref{qmean}).  Thus, our results do not necessarily contradict those of \citet{li98}, who found that the dust models that best fitted 
the observed continuum and $10~\mu$m silicate emission from the disk are those that feature extremely porous ($> 95$\%) grains.  It remains 
to be seen, however, whether a distribution of grain porosities can simultaneously satisfy the constraints provided by scattered-light and 
thermal images of the inner disk.

\subsubsection{Dust Grains beyond the Ice-Sublimation Zone\label{outerdisk}}

Our estimate of $a_{\rm min} = 0.15$--$0.20~\mu$m within 120~AU matches that obtained by \citet{vos99} using ground-based observations 
of the disk's colors and polarization beyond 115~AU \citep{par87,lec93,wol95} and a dust model featuring an $R$-band refractive index 
of $m_R = 1.152 - 0.005i$ and a grain-size distribution of $dn \propto a^{-3.2}~da$.  This refractive index is descriptive of both ``dirty-ice'' 
grains with 50\% porosity and astrosil grains of 76\% porosity.  However, our HRC images show that the disk steadily reddens beyond 120~AU,
so the other dust models considered by \citet{vos99} -- and rejected because they produced colors that were too red compared with the 
nearly neutral colors observed from the ground -- are in fact potentially viable models for this region of the disk.  These models feature
larger refractive indices (corresponding to various combinations of compact and moderately porous grains of astrosil, mixed or layered with 
dirty ice) and larger $a_{\rm min}$.  In another study of the disk's polarization and colors, \citet{kri00} also favored models with 
moderate refractive indices, porosities $\lesssim 50$\%, and depleted numbers of grains smaller than 2--$3~\mu$m.  

Although tabulated values of $Q_{\rm sca}$ and $g$ have not yet been published for moderately porous astrosil grains with icy mantles, it 
is reasonable to expect that the optical colors of an edge-on disk of such grains vary with $a_{\rm min}$ in a manner similar to those
depicted in Figures~\ref{dcolors}a--d.  If so, then the increasingly red color gradient observed in $\beta$~Pic's disk beyond 120~AU 
(Figure~\ref{colors}) may reflect values of $a_{\rm min}$ that increase with distance from $\sim 0.15~\mu$m to perhaps $\sim 2~\mu$m at 
250~AU \citep{vos99}.  A similar trend was proposed by \citet{wei03}, based on the disappearance of the $10~\mu$m silicate emission feature
beyond 20~AU of the star.  These phenomena are consistent with the scenario of a diminished presence of small ($a \lesssim 2$--$3~\mu$m)
grains near the ice-sublimation zone ($\gtrsim 100$~AU), where cometary activity ceases \citep{li98} and radiation pressure sets 
$a_{\rm min}$ to 1--$10~\mu$m, depending on porosity \citep{art88}.  The steeper color gradients in the southwest extension of the disk 
suggest that $a_{\rm min}$ increases more rapidly in that extension, i.e., the number of small grains in the southwest extension is smaller 
than in the northeast extension.  This suggestion is consistent with the observed asymmetry between the polarizations of the two extensions 
\citep{wol95}, which has been attributed to a 20-30\% larger population of small grains in the northeast extension \citep{kri00}.

\subsubsection{Dust Grains in the Secondary Disk\label{secdust}}

The largely uncertain colors of the secondary disk (\S\ref{componentcolor}) prevent us from rigorously comparing the grain characteristics 
of the main and secondary disks.  However, the lack of inflections in the secondary disk's color and surface-brightness profiles 
(Figure~\ref{mainsecprof}) indicates that the grain populations of the two component disks are fundamentally different at 80--150~AU from the 
star.  If the inflections in the main disk's profiles at $\sim 115$~AU are solely caused by ice sublimation, then the composition and/or 
size distribution of the secondary disk's grains must be sufficiently different from those of the main disk that the ice-sublimation 
boundary in the secondary disk (if it exists) lies within or beyond the 80--150~AU range of our analysis.  More precise photometry of the 
secondary disk is needed to determine whether the differences in the grains' composition or size distribution required for such a boundary 
shift are compatible with the colors of the disk.  

\citet{art97} argued that icy grains should not exist anywhere in $\beta$~Pic's disk because they quickly photoevaporate in the face of the 
star's strong ultraviolet flux and because they are structurally brittle and unable to survive high-velocity collisions with other grains.
If so, then other causes of the differences between the color and surface-brightness profiles of the main and secondary disks must be 
considered.  For example, the inflections in the main disk's profiles at $\sim 115$~AU may reflect a sharp decrease in the number density of 
grains within that distance from the star.  \citet{lec96} successfully modeled the inflection in the surface-brightness profile by imposing 
a lower limit on the sizes of comets or asteroids that travel from outer regions of the disk and retain enough volatile material to evaporate 
within 110~AU.  If this scenario is correct, then the lack of inflections in the secondary disk's profiles suggests that most of its dust is 
produced by colliding or evaporating planetesimals orbiting near the star rather than by evaporating comets with large, eccentric orbits.  
This interpretation is consistent with the hypothesis that the ``warp'' in the composite disk is caused by radiation-blown dust from colliding
planetesimals that have been perturbed from the the innermost part of the main disk by a giant planet in an inclined orbit \citep{mou97b,aug01}.

\subsubsection{Possible Effect of Irradiation on Disk Colors\label{colrad}}

Recent photometric studies of Kuiper Belt Objects (KBOs) reveal that these objects exhibit a broad range of optical and near-infrared colors
\citep{luu96,jew98,jew01,teg98,teg03,bar01,del04}.  Growing (but controversial) evidence suggests that KBOs are divided by their perihelia,
$q$, into two color populations: KBOs with $q < 40$~AU have neutral-to-red intrinsic colors, and KBOs with $q > 40$~AU have only red intrinsic 
colors \citep[$0.5 \lesssim B$--$R \lesssim 0.75$, relative to the Sun;][]{teg98,teg03,del04}.  A possible cause of this color trend is the 
diminution of collisions and/or cometary activity in the outer Kuiper Belt \citep{luu96,ste02,del04}.  This hypothesis rests on the notion 
that the ``dirty-ice'' surfaces of KBOs are progressively polymerized and reddened by high-energy solar and cosmic radiation unless they are 
recoated with primordial dust ejected by collisions between KBOs or the sublimation of H$_2$O, CO$_2$, and CO ices.

The observed reddening of $\beta$~Pic's disk beyond its ice-sublimation zone suggests a possible connection with the reddening of distant KBOs.
Could the reddening of $\beta$~Pic's disk beyond $\sim 115$~AU be the result of irradiative polymerization rather than an increasing minimum 
grain size?  \citet{luu96} simulated the rate of irradiative reddening of KBOs using an inverse exponential function with an $e$-folding time 
of $\sim 10^8$~yr.  Because the radiation is overwhelmingly solar within the heliopause \citep{gil02}, we crudely apply this function to the 
region of $\beta$~Pic's disk in our FOV by assuming that the $e$-folding time scales with luminosity and the distance from the star.  In this
situation, the $B$--$V$ color of long-lived icy grains orbiting $\beta$~Pic \citep[$L/L_{\odot} = 8.7$;][]{cri97} would increase by $\sim 0.3$~mag
in a few times $10^8$~yr.  The lifetimes of grains not quickly expelled from the disk by radiation pressure are limited by either 
Poynting--Robertson drag or intergrain collisions, depending on their size and location within the disk.  At 120--250~AU, the minimum lifetimes of 
grains larger than $\sim 0.2~\mu$m are constrained by collisions to $\sim 0.2$--2~Myr \citep{bac93}.  Thus, the lifetimes of the grains observed in 
our HRC images are much too short for the observed color gradient to be caused by irradiative reddening.  This condition likely persists throughout 
the outer disk, because the grain lifetimes beyond $\sim 500$~AU (which are limited by Poynting--Robertson drag) are $\sim 1$\% of the irradiative 
$e$-folding times at those distances.

\subsubsection{Comparing the disks around $\beta$~Pic and AU Mic\label{aumic}}

\begin{figure*}[t]
   \epsscale{1.00}\plotone{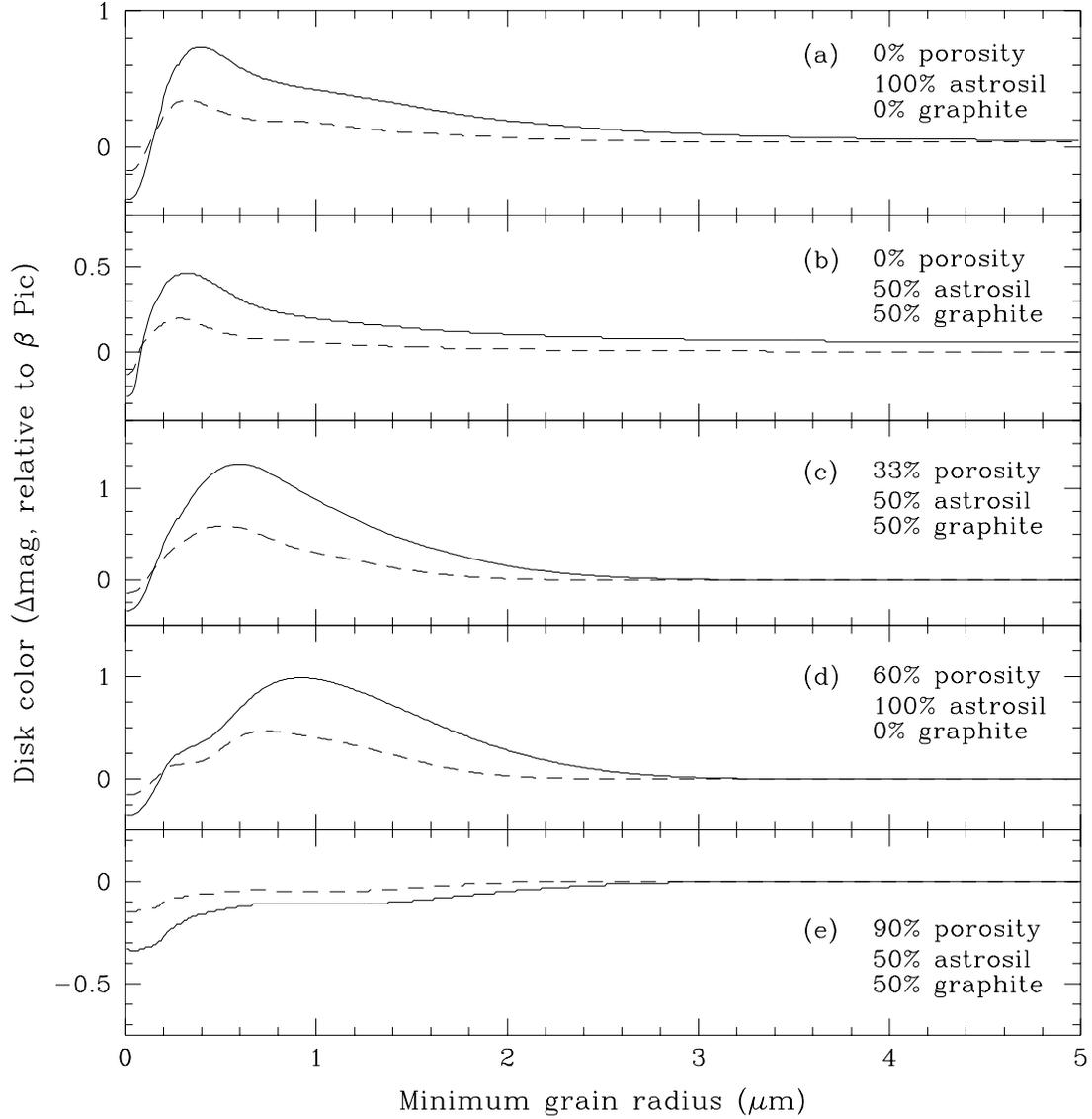}
   \caption{
	Simulated ACS F435W--F606W colors (dashed curves) and F435W--F814W colors (solid curves) of the disk (relative to those of 
	$\beta$~Pic) as functions of minimum grain size for various grain compositions and porosities.  The colors are derived from the 
	corresponding values of $\langle Q_{\rm eff}\rangle$ shown in Figure~\ref{qmean} and a grain-size distribution of $dn \propto 
	a^{-3.5}~da$ \citep{doh69}.
	They pertain to the non-icy grains expected within the ice sublimation zone within $\sim 100$~AU of the star \citep{pan97,li98}.  
	The colors of the composite disk in this region, measured from our PSF-deconvolved images (Figure~\ref{colors}), are 
	F435W--F606W~$\approx 0.1$ and F435W--F814W~$\approx 0.2$.
        }
  \label{dcolors}
\end{figure*}

The disk around the M dwarf AU Mic, whose spatial velocity and age are similar to those of $\beta$~Pic \citep{bar99}, is the only known debris 
disk exhibiting blue colors at optical wavelengths \citep{kri05a}.  Moreover, the disk's F435W--F814W color becomes bluer with increasing 
distance from the star.  The disk extends at least 210~AU from the star and is viewed almost edge-on \citep{kal04}, so the predicted F435W--F814W
colors shown in Figure~\ref{dcolors} can be compared to the observed ACS colors of the disk if the grains are not icy.  In this case, the 
measured F435W--F814W~$= -0.3$ at 30~AU \citep{kri05a} from AU~Mic suggests that the disk may be composed of astrosil and/or graphite grains with 
$a_{\rm min} < 0.1~\mu$m for porosities $\lesssim 60$\% or $a_{\rm min} \approx 0.2~\mu$m for 90\% porosity.  The possibility that AU~Mic's 
disk contains smaller or more porous grains than are evident in $\beta$~Pic's inner disk (\S\ref{innerdisk}) is consistent with the much 
lower radiation pressure exerted on such grains by AU~Mic than by $\beta$~Pic \citep{kal04,kri05a}.  On the other hand, the measured 
F435W--F814W~$= -0.5$ at 60~AU from the star cannot be produced by any combination of grain composition and porosity shown in Figure~\ref{dcolors}, 
so the grains in AU~Mic's disk may have a different mineral composition altogether.  

\section{Summary and Concluding Remarks\label{summary}}

We have presented $B$-, Broad $V$-, and Broad $I$-band coronagraphic images of the dusty debris disk around $\beta$~Pictoris obtained with the High
Resolution Channel of {\it HST's} Advanced Camera for Surveys.  We have exploited the HRC's image resolution and stability by subtracting 
a well-matched coronagraphic reference image and deconvolving the instrumental PSF from the multiband images.  The resultant images provide
the most morphologically detailed and photometrically accurate views of the disk between 30 and 300~AU from the star obtained to date.

Our PSF-deconvolved images confirm that the apparent warp in the disk $\lesssim 100$~AU from the star, which was previously observed by 
\citet{bur95} and \citet{hea00} and modelled by \citet{mou97b} and \citet{aug01}, is a distinct secondary disk inclined to the main disk by 
$\sim 5^{\circ}$.  The opposing extensions of the secondary disk are not collinear, but their outwardly projected midplanes (spines) are 
coincident with the isophotal inflections previously seen at large distances and commonly called the ``butterfly asymmetry'' \citep{kal95}.  
The surface brightness profiles along the spines of the secondary disk from 80 to 150~AU can be fit with single power-law functions having 
indices of $-3.7 < -\alpha < -5.0$.  The lack of inflections in the surface brightness profiles around the expected ice-sublimation boundary 
($\sim 100$~AU) suggests that the composition and/or size distribution of grains in the secondary disk is different from those of the main disk.
Altogether, these phenomena support the notion that the secondary disk and the butterfly asymmetry comprise radiatively expelled dust from 
colliding planetesimals in inclined orbits near the star rather than from evaporating comets in large, eccentric orbits.

We confirm the ``wing-tilt asymmetry'' between the opposing extensions of the main disk, but we find that the effect is centered on 
the star rather than offset toward the southwest extension \citep{kal95}.  While the spine of the northeast extension appears linear 
80--250~AU from the star, the southwest spine is distinctly bowed with an amplitude of $\sim 1$~AU.  The vertical width of 
the main disk within $\sim 120$~AU is nearly constant and is up to 50\% narrower than previously reported.  The clumpy structures 
observed in mid-infrared images and spectra of the disk \citep{wah03,wei03,oka04,tel05} are not seen in our optical scattered-light 
images.  The surface-brightness profiles along the spines of the main disk's extensions can be fit by four distinct power laws separated 
by inflections at $\sim 70$, 117, and 193~AU.  The power-law indices, $\alpha$, match those of \citet{hea00} well within 70~AU of 
$\beta$~Pic, but are 10--15\% smaller than those of \citet{hea00} at 127--193~AU.  This discrepancy cannot be attributed to differences
in the instrumental PSFs.  The indices within 150~AU change significantly after removing the contribution from the secondary disk, so 
the superposed effects of both disks must be considered in future models of $\beta$~Pic's circumstellar dust distribution.

The two-dimensional F606W/F435W and F814W/F435W flux ratios of the composite disk are nonuniform and asymmetric about the projected major and minor 
axes of the disk.  Within 150~AU of $\beta$~Pic, the ratios on the southeast side of the disk are $\sim 1.0$--1.1 times those of the star, 
while those on the northwest side are $\sim 1.2$--1.4 times those of the star.  The redder appearance of the nearer northwest side of the
disk is inconsistent with the expectation that forward scattering from disk grains should diminish with increasing wavelength.  The F435W--F606W
and F435W--F814W colors along the spine of the disk are $\sim 0.1$~mag and $\sim 0.2$~mag redder, respectively, than those of $\beta$~Pic 
at $\sim 40$--120~AU from the star.  These color excesses increase steadily beyond $\sim 120$~AU, respectively reaching $\sim 0.2$--0.35~mag and 
$\sim 0.25$--0.4~mag at 250~AU.  These results contradict the longstanding notion that the disk consists of neutrally scattering grains with 
sizes larger than several $\mu$m \citep{par87,lec93}.

We have compared the colors measured along the spine of the composite disk with those expected for non-icy grains having a number density 
$\propto r^{-3}$ and different compositions, 
porosities, and minimum grain sizes.  We find that the observed F435W--F606W and F435W--F814W colors within the ice-sublimation zone 
($\lesssim 100$~AU) are consistent with those of compact or moderately porous ($P \lesssim 60$\%) grains of astronomical silicate and/or graphite 
with minimum sizes of $\sim 0.15$--$0.20~\mu$m.  The observed colors are inconsistent with the blue colors expected from the very porous grains 
($P \gtrsim 90$\%) that best reproduce the $10~\mu$m silicate emission feature observed within $\sim 35$~AU of the star \citep{li98}.  The 
reddening colors beyond $\sim 120$~AU may reflect the formation of ``dirty ice'' grains or an increasing minimum grain size beyond the 
ice-sublimation boundary.  The latter condition is consistent with the decreased production of submicron grains as cometary activity diminishes.
It is unlikely that the reddening of disk beyond $\sim 120$~AU is caused by irradiative polymerization (as has been postulated for the reddest 
and most distant Kuiper Belt Objects) because the required irradiation time is hundreds of times longer than the expected lifetimes of the grains 
at those distances.

Our ACS/HRC coronagraphic images of $\beta$~Pic's disk are the finest multiband, scattered-light images of its inner region 
(30--300~AU) ever recorded.  These images will not be superseded by the {\it Terrestrial Planet Finder} or other proposed extrasolar-planet
imaging missions because the fields of view of those missions will lie well within the region obscured by the HRC's occulting spot.  Comparable
infrared images of the disk are expected from the {\it James Webb Space Telescope} if the current specifications for its coronagraphic-imaging 
modes are maintained.  Thus, our observations and derived results should be standard references for comparative and theoretical studies of 
circumstellar debris disks for at least the next decade.

\acknowledgments 
We thank N.~Voshchinnikov for his advice regarding the optical properties of porous grains.  We also thank E.~Pantin for his comments on the 
manuscript.  ACS was developed under NASA contract NAS~5-32865, and this research has been supported by NASA grant NAG5-7697 and by an equipment
grant from  Sun Microsystems, Inc.  The STScI is operated by AURA Inc., under NASA contract NAS5-26555.

\clearpage 

\begin{deluxetable}{lcclrcl}
\tabletypesize{\footnotesize}
\tablewidth{270pt}
\tablecaption{Log of ACS/HRC exposures\tablenotemark{a}\label{explog}}
\tablehead{
               & \colhead{Roll angle\tablenotemark{b}} &                 &                 & \multicolumn{3}{c}{Exposures} \\
\cline{5-7}
\colhead{Star} & \colhead{(deg)}                       & \colhead{Orbit} &\colhead{Filter} & \colhead{No.} & \colhead{$\times$} & \colhead{Time (s)} 
}
\tablecolumns{7}
\startdata 
$\alpha$~Pic    & ~87.2                                & 1               & F435W            & 1 & $\times$ & 40  \\
                &                                      &                 &                  & 2 & $\times$ & 100 \\
                &                                      &                 &                  & 2 & $\times$ & 225 \\
                &                                      &                 & F606W            & 1 & $\times$ & 15  \\
                &                                      &                 &                  & 2 & $\times$ & 50  \\
                &                                      &                 &                  & 2 & $\times$ & 200 \\
                &                                      &                 & F814W            & 1 & $\times$ & 40  \\
                &                                      &                 &                  & 2 & $\times$ & 100 \\
                &                                      &                 &                  & 2 & $\times$ & 200 \\
$\beta$~Pic     & 110.0                                & 2               & F606W            & 1 & $\times$ & 10  \\
                &                                      &                 &                  & 2 & $\times$ & 100 \\
                &                                      &                 &                  & 6 & $\times$ & 338 \\
                &                                      & 3               & F435W            & 1 & $\times$ & 10  \\
                &                                      &                 &                  & 2 & $\times$ & 250 \\
                &                                      &                 &                  & 3 & $\times$ & 653 \\
                &                                      & 4               & F814W            & 1 & $\times$ & 10  \\
                &                                      &                 &                  & 2 & $\times$ & 250 \\
                &                                      &                 &                  & 3 & $\times$ & 645 \\
$\beta$~Pic     & 100.3                                & 5               & F606W            & 1 & $\times$ & 10  \\
                &                                      &                 &                  & 2 & $\times$ & 100 \\
                &                                      &                 &                  & 6 & $\times$ & 338 \\
                &                                      & 6               & F435W            & 1 & $\times$ & 10  \\
                &                                      &                 &                  & 2 & $\times$ & 250 \\
                &                                      &                 &                  & 3 & $\times$ & 653 \\
                &                                      & 7               & F814W            & 1 & $\times$ & 10  \\
                &                                      &                 &                  & 2 & $\times$ & 250 \\
                &                                      &                 &                  & 3 & $\times$ & 645 
\enddata

\tablenotetext{a}{\footnotesize Exposures recorded over seven consecutive {\it HST} orbits on UT~2003 October~1.}
\tablenotetext{b}{\footnotesize Defined as the position angle of the $y$-axis of the raw HRC image, measured east of north.}

\end{deluxetable}

\newpage
\begin{deluxetable}{lcrrrr}
\tablewidth{305pt}
\tablecaption{Relative position angles of secondary disk\tablenotemark{a}\label{tiltang}}
\tablehead{
\colhead{Extension} & 		      & \colhead{F435W} & \colhead{F606W} & \colhead{F814W} & Average 
}
\tablecolumns{6}
\startdata
Northeast       & &  $6^{\circ}\!{.}3$ &  $5^{\circ}\!{.}3$ &  $6^{\circ}\!{.}3$ &  $5^{\circ}\!{.}9 \pm 0^{\circ}\!{.}6$ \\
Southwest       & & $-4^{\circ}\!{.}4$ & $-4^{\circ}\!{.}9$ & $-4^{\circ}\!{.}8$ & $~-4^{\circ}\!{.}7 \pm 0^{\circ}\!{.}3$ 
\enddata

\tablenotetext{a}{\footnotesize The position angles pertain to regions of the secondary disk between the vertical dotted lines 
in Figure~\ref{traceprof}, and are measured relative to the corresponding extensions of the main disk.}

\end{deluxetable}

\clearpage
\begin{deluxetable}{lccllll}
\tabletypesize{\footnotesize}
\tablewidth{365pt}
\tablecaption{Power-law fits to surface-brightness profiles of composite disk\label{powlaw1}}
\tablehead{
                    & \colhead{PSF}    & & \hspace*{0.1in}  & \multicolumn{3}{c}{Power-law index ($\alpha$)\tablenotemark{a}} \\
\cline{5-7}
\colhead{Extension} & \colhead{Decon?} & \colhead{Region\tablenotemark{b}} & & \colhead{F435W}    & \colhead{F606W}    & \colhead{F814W}
}
\tablecolumns{7}
\startdata
Northeast           & No                   & 1 & & $1.34 \pm 0.02$ & $1.34 \pm 0.02$ & $1.30 \pm 0.02$ \\
                    &                      & 2 & & $1.91 \pm 0.02$ & $1.91 \pm 0.02$ & $1.90 \pm 0.02$ \\
                    &                      & 3 & & $4.28 \pm 0.02$ & $4.19 \pm 0.02$ & $4.04 \pm 0.02$ \\
                    &                      & 4 & & $3.51 \pm 0.07$ & $3.63 \pm 0.04$ & $3.49 \pm 0.06$ \\
\cline{2-7}
                    & Yes		   & 1 & & $1.45 \pm 0.04$ & $1.43 \pm 0.03$ & $1.38 \pm 0.03$ \\
                    &                      & 2 & & $1.84 \pm 0.04$ & $1.83 \pm 0.03$ & $1.80 \pm 0.03$ \\
                    &                      & 3 & & $4.20 \pm 0.05$ & $4.19 \pm 0.04$ & $4.12 \pm 0.04$ \\
                    &                      & 4\tablenotemark{c} & & $2.80 \pm 0.15$ & $3.34 \pm 0.11$ & $3.34 \pm 0.12$ \\
\cline{1-7}
Southwest           & No                   & 1 & & $1.02 \pm 0.02$ & $1.06 \pm 0.02$ & $1.15 \pm 0.02$ \\
                    &                      & 2 & & $2.03 \pm 0.02$ & $2.03 \pm 0.02$ & $2.05 \pm 0.02$ \\
                    &                      & 3 & & $4.89 \pm 0.03$ & $4.76 \pm 0.02$ & $4.56 \pm 0.02$ \\
                    &                      & 4 & & $3.99 \pm 0.09$ & $4.00 \pm 0.05$ & $3.84 \pm 0.07$ \\
\cline{2-7}
                    & Yes		   & 1 & & $1.08 \pm 0.04$ & $1.10 \pm 0.03$ & $1.21 \pm 0.04$ \\
                    &                      & 2 & & $1.93 \pm 0.04$ & $1.94 \pm 0.03$ & $1.97 \pm 0.03$ \\
                    &                      & 3 & & $4.82 \pm 0.06$ & $4.75 \pm 0.05$ & $4.60 \pm 0.05$ \\
                    &                      & 4\tablenotemark{c} & & $3.11 \pm 0.18$ & $3.53 \pm 0.12$ & $3.26 \pm 0.14$
\enddata

\tablenotetext{a}{\footnotesize $\alpha$ is the index of the radial power-law, $r^{-\alpha}$, that best fits the surface brightness
of the spine within a given region of each disk extension.  Equivalently, $-\alpha$ is the slope of the linear least-squares fit to 
the given region of the logarithm of the profile, as shown in Figure~\ref{midpfit}.}
\tablenotetext{b}{\footnotesize Angular and projected distances of regions from $\beta$~Pic: (1) 2\farcs0--3\farcs5 (39--67 AU); 
(2) 3\farcs7--5\farcs6 (71--108 AU); (3) 6\farcs6--10\farcs0 (127--193 AU); (4) 10\farcs0--13\farcs4 (193--258 AU).}
\tablenotetext{c}{\footnotesize Indices in region~4 of the PSF-deconvolved images are suspect because the efficacy of the 
Lucy--Richardson algorithm is diminished in regions of low S/N.}

\end{deluxetable}

\begin{deluxetable}{llllrllll}
\tabletypesize{\footnotesize}
\tablewidth{430pt}
\tablecaption{Power-law fits to surface-brightness profiles of component disks\tablenotemark{a}\label{powlaw2}}
\tablehead{
                    & & \hspace*{0.1in} & \multicolumn{2}{c}{Region} & \hspace*{0.1in} & \multicolumn{3}{c}{Power-law index ($\alpha$)\tablenotemark{b}} \\
\cline{4-5}\cline{7-9}
\colhead{Extension} & \colhead{Component} & & \colhead{(arcsec)} & \colhead{(AU)} & & \colhead{F435W} & \colhead{F606W} & \colhead{F814W}
}
\tablecolumns{9}
\startdata
Northeast           & Main      & & 4.1--5.6 & 80--108  & & $1.23 \pm 0.06$ & $1.42 \pm 0.07$ & $1.19 \pm 0.08$ \\
                    & Main      & & 6.6--7.8 & 127--150 & & $5.09 \pm 0.12$ & $4.52 \pm 0.11$ & $4.38 \pm 0.25$ \\
                    & Secondary & & 4.1--7.8 & 80--150  & & $4.46 \pm 0.14$ & $4.63 \pm 0.14$ & $4.97 \pm 0.24$ \\
& \\
Southwest           & Main      & & 4.1--5.6 & 80--108  & & $1.50 \pm 0.05$ & $1.64 \pm 0.06$ & $1.58 \pm 0.06$ \\
                    & Main      & & 6.6--7.8 & 127--150 & & $5.23 \pm 0.16$ & $5.18 \pm 0.14$ & $4.96 \pm 0.12$ \\
                    & Secondary & & 4.1--7.8 & 80--150  & & $3.80 \pm 0.10$ & $3.72 \pm 0.13$ & $3.93 \pm 0.10$
\enddata

\tablenotetext{a}{\footnotesize Profiles are extracted from the PSF-deconvolved images and shown in Figure~\ref{mainsecprof}.}
\tablenotetext{b}{\footnotesize See note (a) of Table~\ref{powlaw1}.}

\end{deluxetable}

\newpage

\clearpage

\clearpage

\clearpage

\clearpage

\clearpage

\clearpage

\clearpage

\clearpage

\clearpage

\clearpage

\clearpage

\clearpage

\clearpage

\clearpage

\clearpage

\clearpage

\clearpage

\clearpage

\clearpage

\clearpage

\end{document}